\documentclass[a4paper,11pt]{article}
\usepackage{fullpage}
\usepackage{authblk}
\usepackage[urlcolor=blue,citecolor=black,linkcolor=black]{hyperref}
\usepackage{natbib}
\usepackage{graphicx}

\title{On computing viscoelastic Love numbers for general planetary models: the \texttt{ALMA${}^3$} code}

\author[1]{Daniele Melini}
\author[2]{Christelle Saliby}
\author[3]{Giorgio Spada}
\affil[1]{Istituto Nazionale di Geofisica e Vulcanologia, Roma, Italy}
\affil[2]{G\'eoazur, CNRS, Observatoire de la C\^{o}te~d'Azur, Universit\'e~C\^{o}te~d'Azur, Valbonne, France}
\affil[3]{Dipartimento di Fisica e Astronomia (DIFA), Alma Mater Studiorum Universit\`a di Bologna, Italy}
\date{}

\newcommand{\alma}{\texttt{ALMA}${}^3$}
\newcommand{\binom}[2]{ {#1 \choose #2} }

\begin{document}

\maketitle
\linespread{1.4}

\textit{This is a pre-copyedited, author-produced PDF of an article accepted for publication in Geophysical Journal International following peer review. The version of record [D. Melini, C. Saliby, G. Spada, On computing viscoelastic Love numbers for general planetary models: the ALMA${}^3$ code, Geophysical Journal International, Volume 231, Issue 3, December 2022, Pages 1502–1517] is available online at: }\url{https://doi.org/10.1093/gji/ggac263}

\begin{abstract}
{The computation of the Love numbers for a spherically symmetric self-gravitating
viscoelastic Earth {is a classical problem} in global geodynamics}. 
{Here we revisit the problem of the numerical evaluation of loading and tidal Love numbers in the static limit for an incompressible planetary body, adopting a Laplace inversion scheme based upon the Post-Widder formula as an alternative to the {traditional viscoelastic} normal modes method.} {We also consider, {whithin the same framework,} complex-valued, frequency-dependent Love numbers that describe the response to a periodic forcing, which are paramount in the study of the tidal deformation of planets. Furthermore, we numerically obtain the time-derivatives of Love numbers, suitable for modeling geodetic signals in response to surface loads variations}. {A number of examples are shown, in which time and frequency-dependent Love numbers are evaluated for the Earth and planets adopting realistic rheological profiles.} {The numerical solution scheme is implemented in \alma{}} (the plAnetary Love nuMbers cAlculator, version 3), an {upgraded} open-source Fortran 90 {program} that computes {the} Love numbers for radially layered planetary bodies with a wide range of rheologies, including transient laws like Andrade or Burgers. 


\end{abstract}

\section{Introduction}

Love numbers, first introduced by A.E.H. Love in 1911, provide a complete description of the response of a planetary body to external, surface or internal perturbations. In his seminal work, \citet{love1911some} defined the Love numbers (LN) in the context of computing the radial deformation and the perturbation of gravity potential for an elastic, self-gravitating, homogeneous sphere that is subject to the gravitational pull of a tide-raising body. This definition has been subsequently extended by \citet{shida1912elasticity} to include also horizontal displacements. In order to describe the response to surface loads, an additional set of LNs, dubbed \emph{loading Love numbers}, has been introduced in order to describe the Earth's response to surface loads 
\citep[see \emph{e.g.,}][]{munk1960rotation,farrell1972deformation} and today they are routinely used in the 
context of the Post Glacial Rebound problem \citep{spada2011benchmark}. In a similar way, \emph{shear Love numbers} represent the response to a shear stress acting on the surface \citep{saito1978relationship} while \emph{dislocation Love numbers} describe deformations induced by internal point dislocations \citep[see \emph{e.g.},][]{sunokubo1993internaldislocation}.

The LN formalism has been originally defined in the realm of purely elastic deformations, for spherically symmetric Earth models consistent with global seismological observations. 
However, invoking the Correspondence Principle in linear viscoelasticity~\citep[see \emph{e.g.,}][]{christensen1982}, the LNs can be generalized to anelastic models in a straightforward way. Currently, viscoelastic LNs are a key ingredient of several geophysical applications involving the time-dependent response of a {spherically symmetric Earth model} to surface loads or endogenous perturbations. For example, they are essential to the solution of the Sea Level Equation \citep{farrell76on} and are exploited in current numerical implementations of the Glacial Isostatic Adjustment (GIA) problem, either on  millennial~\citep[see \textit{e.g.},][]{spada2019selen4-gmd-12-5055-2019} or on decadal time scale \citep[see \textit{e.g.},][]{Melini15on}.

Since LNs depend on the internal structure of a planet and on its constitution, they can provide a means of establishing constraints on some physical parameters of the planet interior on the basis of geodetic measurements or astronomic observations \citep[see \textit{e.g.},][]{zhang1992lovenumbers, kellerman2018exoplanets}. For tidal periodic perturbations, complex LNs can be defined in the frequency domain, accounting for both the amplitude and phase lag of the response to a given tidal frequency \citep{williams2015moon}. Frequency-domain LNs are widely used to constrain the interior structure of planetary bodies on the basis of observations of {tidal amplitude and phase lag} \citep[see \emph{e.g.,}][]{sohl2003titan,dumoulin2017tidal,tobie2019exoplanets}, to study the state of stress of satellites induced by tidal {forcings} \citep[see \emph{e.g.,}][]{wahr2009modeling} or to investigate the {tidal} response of {the} giant planets \citep[see \emph{e.g.,}][]{gavrilov1977love}.

Viscoelastic LNs for a spherically symmetric, radially layered, self-gravitating planet are traditionally computed within the framework of the ``viscoelastic normal modes'' method {introduced by}~\citet{peltier1974the}, which relies upon the solution of Laplace-transformed equilibrium equations using the formalism of elastic propagators. As discussed \emph{e.g.} by \citet{spadaboschi2006using} and \citet{melini2008post}, this approach becomes progressively less feasible as the detail of the rheological model is increased or if complex constitutive laws are considered. Several workarounds have been proposed in the literature to avoid these shortcomings \citep[see, \emph{e.g.}][]{rundle1982viscoelastic,friederich1995complete,riva2002approximation,tanaka2006new}. Among these, the Post-Widder Laplace inversion formula \citep{post1930generalized,widder1934inversion}, first applied by \citet{spadaboschi2006using} to the evaluation of viscoelastic LNs for the Earth, has the advantage of maintaining unaltered the formal structure of the viscoelastic normal modes and {of allowing 
for} a straightforward implementation of complex rheological laws. {For periodic loads, alternative numerical integration schemes similar to those developed by \citet{takeuchi1972seismic} for the elastic problem \citep{na2011computation,wang2012load} have been applied to the viscoelastic case by integrating Fourier-transformed solutions \citep{tobie2005tidal,tobie2019exoplanets}. }

In this work, we revisit the Post-Widder approach to the evaluation of LNs with the aim of extending it to more general planetary models, relaxing {some} of the assumptions originally made by \citet{spadaboschi2006using}.
{In particular, we introduce a {layered} core in the Post-Widder formalism and obtain analytical expressions}
for the time derivatives of LNs, needed to model geodetic velocities in response to the variation of surface loads.
In this respect, our approach is complementary to that of \citet{padovan2018matrix}, who derived a semi-analytical solution for the fluid LNs using the propagator formalism.
We implement our results in \alma~ (the plAnetary Love nuMbers cAlculator, version 3), an open-source code which extends and generalizes the program originally released by \citet{spada2008alma}.  {\alma{} introduces a range of new capabilities, including the evaluation of  frequency-domain LNs describing the response to periodic forcings,  suitable for studying tidal dissipation in the Earth and planets.}

This paper is organized as follows. In Section  \ref{sect:theory} we give a brief outline of the theory underlying the computation of viscoelastic LNs and of the application of the Post-Widder Laplace inversion formula. 
In Section \ref{sect:alma} we discuss some general aspects of \alma, leaving the technical details to a User Manual. In Section \ref{sect:benchmark} {we validate \alma~ through some benchmarks between our numerical results and available reference solutions} 
In Section \ref{sect:examples} we discuss some numerical examples before drawing our conclusions in Section \ref{sec:conclusions}. 

\section{Mathematical background}\label{sect:theory}

The details of the Post-Widder approach to numerical Laplace inversion have been extensively discussed in previous works \citep[see][]{spadaboschi2006using,spada2008alma,melini2008post}. 
{In what follows, we {only} give a brief account of the Post-Widder Laplace inversion method for the sake of illustrating how the new features of \alma~ have been implemented within its context.}

\subsection{Viscoelastic normal modes}

Closed-form analytical expressions for the LNs exist only for a few extremely simplified planetary models. The first is the homogeneous, self-gravitating sphere, often referred to as the ``Kelvin sphere''
\citep[][]{thomson1863xxvii}. The second is the two-layer, incompressible, non self-gravitating model that has been solved analytically by \citet{wu1996some}. For more complex models, LNs shall be computed either through fully numerical integration of the equilibrium equations, or by invoking semi-analytical schemes. Among the latter, the viscoelastic normal modes method, introduced by \citet{peltier1974the}, relies upon the solution of the equilibrium equations in the Laplace-transformed domain. Invoking the {Correspondence Principle} {\citep[\emph{e.g.},][]{christensen1982}} the equilibrium equations can be cast in a formally elastic form by defining a complex rigidity $\mu(s)$ that depends on the rheology adopted and is a function of the Laplace variable $s$. 

Following \citet{spadaboschi2006using}, at a given harmonic degree $n$, the Laplace-transformed equations can be solved with standard propagator methods, and their solution at the planet surface ($r=a$) can be written in vector form as
\begin{equation}\label{eq:xsol}
    \tilde{\mathbf{x}}(s) = \tilde{f}(s) \Big(P_1\Lambda(s)J\Big)  \Big(P_2\Lambda(s)J\Big)^{-1} \mathbf{b}\,,
\end{equation}
where the tilde denotes Laplace-transformed quantities, vector 
$\tilde{\mathbf{x}}(s) = \left( \tilde{u}, \tilde{v}, \tilde{\varphi} \right)^T$
 contains the $n$-th degree harmonic coefficients of the vertical ($\tilde u$) and horizontal ($\tilde v$) components of the displacement field and the incremental potential ($\tilde \varphi$), $\tilde{f}(s)$ is the Laplace-transformed time-history of the  forcing term, $P_1$ and $P_2$ are appropriate $3 \times 6$ projection operators, $J$ is a $6\times 3$ array that accounts for the boundary conditions at the core interface, and $\mathbf{b}$ is a three-component vector expressing the surface boundary conditions (either of loading or of tidal type). 
 In Eq. (\ref{eq:xsol}), $\Lambda(s)$ is a $6\times 6$ array that propagates the solution from the core radius ($r=c$) to the planet surface ($r=a$), which has the form: 
\begin{equation}\label{eq:prop}
\Lambda(s) = \prod_{k=N}^1 Y_k(r_{k+1},s) Y_k^{-1}(r_k,s)\,,
\end{equation}
where $N$ is the number of homogeneous layers outside the planet core, $r_k$ is the radius of the interface between the $(k-1)$-th and $k$-th layer, with $r_1\le ... \le r_N$, $r_1=c$ and $r_{N+1}=a$. In Eq.~(\ref{eq:prop}), $Y_k(r,s)$ is the fundamental matrix that contains the six linearly independent solutions of the equilibrium equations valid in the $k$-th layer, whose expressions are given analytically in~\citet{sabadini1982polar}. When incompressibility is assumed, the matrix $Y_k(r,s)$ depends upon the rheological constitutive law through the functional form of the complex rigidity $\mu(s)$, which replaces the elastic rigidity $\mu$ of the elastic propagator~\citep{wu1982viscous}. Table~\ref{tab:rheologies}\marginpar{T\ref{tab:rheologies}} lists expressions of $\mu(s)$ for some rheological laws.
For a fluid inviscid (\emph{i.e.,} zero viscosity) core, the array $J$ in Eq.~(\ref{eq:xsol}) is a $6\times 3$ interface matrix whose components are explicitly given by \citet{sabadini1982polar}; {conversely, for a solid core, } $J$ corresponds to the $6\times 3$ portion of the fundamental matrix for the core $Y_c(c,s)$ that contains the three solutions behaving  regularly for $r\mapsto 0$.

From the solution $\tilde{\mathbf{x}}(s)$ obtained in (\ref{eq:xsol}), the Laplace-transformed {Love numbers} are defined as:
\begin{eqnarray}
\tilde{h}_n(s) &=& \frac{m}{a} \tilde{u}_n(s) \label{eq:h_s} \\
\tilde{l}_n(s) &=& \frac{m}{a} \tilde{v}_n(s)\label{eq:l_s}\\
\tilde{k}_n(s) &=& -1-\frac{m}{ag} \tilde{\varphi}_n(s) \label{eq:k_s}\,,
\end{eqnarray}
where we have made the $n$-dependence explicit, $m$ is the {mass of the planet} and $g$ is the unperturbed surface gravitational acceleration \citep[][]{farrell1972deformation,wu1982viscous}. Using Cauchy's residue theorem, for Maxwell or generalized Maxwell rheologies Eqs.~(\ref{eq:h_s}-\ref{eq:k_s}) can be cast in the standard normal modes form, which for an impulsive load ($\tilde{f}(s)=1$) reads 
\begin{equation}\label{eq:xns}
\tilde{L}_n(s) =  L_n^e + \sum_{k=1}^{N_M} \frac{L_n^k}{s-s_n^k}  \,,
\end{equation}
where $\tilde{L}_n(s)$ denotes any of the three LNs, $L_n^e$ is the elastic component of the LN (\emph{i.e.,} the limit for $s\mapsto \infty$), $L_n^k$ are the viscoelastic components
(residues), $s_n^k$ are the (real and negative) roots of the secular equation $\textrm{Det}(P_2\Lambda(s)J)=0$, and where $N_M$ is the number of viscoelastic normal modes, each corresponding to one root of the secular equation \citep{spadaboschi2006using}. However, such standard form is not always available, since 
for some particular rheologies the complex rigidity $\mu(s)$ cannot be cast in the form of a rational fraction (this occurs, for example, for the Andrade's rheology, see Table \ref{tab:rheologies}). This is one of the motivations for adopting non-conventional Laplace inversion formulas like the one discussed in next section.

\subsection{Love numbers in the time domain}\label{sec:lntd}

To obtain the time-domain LNs $h_n(t)$, $l_n(t)$ and $k_n(t)$, it is necessary to perform the inverse Laplace transform of Eqs.~(\ref{eq:h_s}-\ref{eq:k_s}). Within the viscoelastic normal-mode approach, this is usually accomplished through an integration over a (modified) Bromwich path in the complex plane, by invoking the residue theorem. 
In this case, the inversion of~Eq.~(\ref{eq:xns}) yields the time-domain Love numbers in the form:
\begin{equation}\label{eq:xnt}
L_n(t) = L_n^e\delta(t) + H(t) \sum_{k=1}^{N_M} {L_n^k} \textrm{e}^{s_n^k \,t} \,,
\end{equation}
where $\delta(t)$ is the Dirac delta and $H(t)$ is the Heaviside step function
defined by Eq.~(\ref{eq:timehist1}) below, and an impulsive time history {is} assumed ($\tilde{f}(s)=1$). 
As discussed by \citet{spadaboschi2006using}, the traditional scheme of the viscoelastic normal modes suffers from a few but significant shortcomings that, with models of increasing complexity, effectively hinders a reliable numerical inverse transformation. Indeed, the application of the residue theorem demands the identification of the poles of the Laplace-transformed solutions (see Eqs.~\ref{eq:h_s}-\ref{eq:k_s}), which are the roots of {the secular  polynomial equation whose algebraic} degree increases with the number of rheologically distinct layers. In addition, its algebraic complexity may be {unpractical} to handle, particularly for constitutive laws characterized by many material parameters.  

As shown by \citet{spadaboschi2006using} and \citet{spada2008alma}, a possible way to circumvent these difficulties is to compute the inverse Laplace transform through the Post-Widder (PW) formula \citep{post1930generalized,widder1934inversion}. 
 We note, however, that other viable possibilities exist, as the one recently discussed by \citet{Michel-and-Boy-2021}, who have employed Fourier techniques to avoid some of the problems inherent in the Laplace transform method. 
 {While Fourier techniques may be more appropriate to {take complex rheologies into account}, and are clearly more relevant to address Love numbers at tidal frequencies, the motivation of our approach is to address in a unified framework the computation of LNs describing both tidal and surface loads.}
If $\tilde{F}(s) = {\cal L}(F(t))$ is the Laplace transform of $F(t)$, the PW formula gives an asymptotic approximation of the inverse Laplace transform ${\cal L}^{-1}(\tilde{F}(s))$ as a function of the $n$-th derivatives of ${\tilde F}(s)$ evaluated along the real positive axis:
\begin{equation}\label{eq:postwid1}
F(t) = \lim_{n\to\infty} \frac{(-1)^n}{n!} 
\left(\frac{n}{t}\right)^{n+1}\left[ \frac{d^n}{ds^n} \tilde{F}(s) \right]_{s=\frac{n}{t}}\,.
\end{equation}

In general, an analytical expression for the $n$-th derivative of $\tilde{F}(s)$ required in Eq.~(\ref{eq:postwid1}) is not available. By employing a recursive discrete approximation of the derivative and rearranging the corresponding terms, \citet{gaver1966observing} has shown that an equivalent expression is 
\begin{equation}\label{eq:postwid2}
F(t) = \lim_{n\to\infty} 
\frac{n \ln 2}{t}\binom{2n}{n}
\sum_{j=0}^n (-1)^j \binom{n}{j} 
\tilde{F}\left(\frac{(n+j)\ln 2}{t}\right)
\,,
\end{equation}
where the inverse transform $F(t)$ is expressed in terms of samples of the Laplace transform $\tilde{F}(s)$ on the real positive axis of the complex plane. Since for a stably stratified incompressible planet all the singularities of $\tilde{\mathbf{x}}(s)$ (Eq.~\ref{eq:xsol}) are expected to be located along the real negative axis that ensures the long-term gravitational stability \citep{vermeersen2000gravitational}, Eq.~(\ref{eq:postwid2}) provides a strategy for evaluating the time-dependent LNs without the numerical complexities associated with the traditional contour integration. However, as discussed by \citet{valko2004comparison}, the numerical convergence of (\ref{eq:postwid2}) is logarithmically slow, and the oscillating terms can lead to catastrophic loss of numerical precision. 
\citet{stehfest1970algorithm} has shown that, for practical applications, the convergence of
Eq.~(\ref{eq:postwid2}) can be accelerated by re-writing it in the form
\begin{equation}\label{eq:postwid3}
F(t) = \lim_{M\to\infty}
\frac{\ln 2}{t} \,\,\sum_{j=1}^{2M}
\zeta_{j,M} \tilde{F}\left(\frac{j \ln 2}{t}\right)\,,
\end{equation}
where $M$ is the order of the Gaver sequence and where the $\zeta$ constants 
{are} 
\begin{equation}\label{eq:postwid4}
\zeta_{k,M} = (-1)^{M+k}\!\!\! \sum_{j=\mathrm{floor}\left(\frac{k+1}{2}\right)}^{\min(M,k)}
\frac{j^{M+1}}{M!} \binom{M}{j}\binom{2j}{j}\binom{j}{k-j}\,,
\end{equation}
with $\mathrm{floor}(x)$ being the greatest integer less or equal to $x$. Eq.~(\ref{eq:postwid3}) can be applied to (\ref{eq:xsol}) to obtain an $M$-th order approximation of the time-domain solution vector:
\begin{equation}\label{eq:xtsol}
\mathbf{x}^{(M)}(t) =
\frac{\ln 2}{t} \,\,\sum_{j=1}^{2M}
\zeta_{j,M} \, \tilde{\mathbf{x}}\left(\frac{j \ln 2}{t}\right)\,,
\end{equation}
from which the time-domain LNs can be readily obtained according to Eqs.~(\ref{eq:h_s}-\ref{eq:k_s}). 

Recalling that the Laplace transform of $F(t)$ and that of its time derivative $\dot{F}(t)$ are related by
${\cal L}(\dot{F}(t)) = s {\cal{L}}(F(t)) - F(0^-)$
and being $\mathbf{x}(t)=0$ for $t<0$, it is also possible to write an asymptotic approximation for the time derivative of the solution:
\begin{equation}\label{eq:xdtsol}
\dot{\mathbf{x}}^{(M)}(t) =
\left(\frac{\ln 2}{t}\right)^2 \,\,\sum_{j=1}^{2M}
j\,\zeta_{j,M}\, \tilde{\mathbf{x}}\left(\frac{j \ln 2}{t}\right)\,,
\end{equation}
from which the time derivative of the LNs $\dot{h}_n(t)$, $\dot{l}_n(t)$ and $\dot{k}_n(t)$ can be obtained according to Eqs.~(\ref{eq:h_s}-\ref{eq:k_s}). 
{The numerical computation of the time-derivatives of the LNs according to Eq.~(\ref{eq:xdtsol}) is one of the new features introduced in \alma{}.}

The time dependence of the solution vector obtained through Eqs. (\ref{eq:xtsol}-\ref{eq:xdtsol}) is also determined by the time history of the forcing term (either of loading or tidal type), whose Laplace transform $\tilde{f}(s)$ appears in Eq.~(\ref{eq:xsol}). If the loading is instantaneously switched on at $t=0$, its time history is represented by the Heaviside {(left-continuous)} step function
\begin{equation}\label{eq:timehist1}
H(t) = \left\{ 
\begin{array}{ll}
0, & t\le 0 \\
1, & t > 0\,,
\end{array}
\right.
\end{equation}
whose Laplace transform is
\begin{equation}\label{eq:timehist2}
\tilde{H}(s)={\cal L}(H(t)) = \frac{1}{s}\,.
\end{equation}
Since any piece-wise constant function can be expressed as a linear combination of shifted Heaviside step functions \citep[see, \emph{e.g.}][]{spada2019selen4-gmd-12-5055-2019}, LNs obtained assuming the loading time history in Eq. (\ref{eq:timehist1}) can be used to compute the response to arbitrary piece-wise constant loads. However, for some applications, it may be more convenient to represent the load time history as a piece-wise linear function. It is easy to show that any such function can be written as a linear combination of shifted \textit{elementary ramp functions} of length $t_r$, 
of the type
\begin{equation}\label{eq:timehist3}
R(t) = \left\{ 
\begin{array}{cl}
0, & t\le 0 \\
\displaystyle \frac{t}{t_r}, & 0 < t \le t_r \\
1, & t > t_r\,,
\end{array}
\right.
\end{equation}
whose Laplace transform is
\begin{equation}\label{eq:timehist4}
\tilde{R}(s)={\cal L}(R(t)) = 
\frac{1}{s} \cdot \frac{1-e^{-s t_r}}{s\, t_r}\,.
\end{equation}

{Laplace-transformed LNs} corresponding to a step-wise or ramp-wise forcing time history can be obtained by setting  $\tilde{f}(s) = \tilde{H}(s)$ or $\tilde{f}(s)=\tilde{R}(s)$ {in Eq.~(\ref{eq:xsol})}. {The ramp-wise
forcing function {defined by Eq.~(\ref{eq:timehist3})} is one of the new features introduced in \alma{}.}

\subsection{Frequency dependent Love numbers}

In the context of planetary tidal deformation, it is important to determine the response to an external periodic tidal potential.
{The previous version of \texttt{ALMA}} was limited to the case of an instantaneously applied forcing. For periodic potentials, the time dependence of the forcing term has the {oscillating} form $e^{i\omega t}$, {where} 
\begin{equation}
{\omega=\frac{2\pi}{T}}    
\end{equation}
is the angular frequency of the forcing term, $T$ is {the} period {of the oscillation} and
$i=\sqrt{-1}$ is the imaginary unit. In the time domain, the solution vector can be {cast in the form}
\begin{equation}\label{eq:lnfrq1}
\mathbf{x}_\omega(t) = \mathbf{x}_\delta(t) * e^{i\omega t}\,,
\end{equation}
where $\mathbf{x}_\delta(t)$ is the time-domain response to an impulsive ($\delta$-like) load {and the asterisk indicates} the time convolution. Since the impulsive load is {a causal function}, $\mathbf{x}_\delta(t)=0$ for $t<0$ and Eq.~(\ref{eq:lnfrq1}) can be {expressed} as
\begin{equation}\label{eq:lnfrq2}
\mathbf{x}_\omega(t) = e^{i\omega t} \int_0^{\infty} \mathbf{x}_\delta(t') e^{-i\omega t'} dt' =  \mathbf{x}_0(\omega) e^{i\omega t}\,,
\end{equation}
where $\mathbf{x}_0(\omega)$ is the Laplace transform of $\mathbf{x}_\delta(t)$ evaluated at $s=i\omega$. By setting $\tilde{f}(s) = {\cal L}(\delta(t))=1$ and $s=i\omega$ in Eq.~(\ref{eq:xsol}), we obtain
\begin{equation}\label{eq:xfreq1}
\mathbf{x}_0(\omega) = 
\Big(P_1\Lambda(i\omega)J\Big)  \Big(P_2\Lambda(i\omega)J\Big)^{-1} \,\mathbf{b}\,.
\end{equation}
{Hence, in} analogy with Eqs.~(\ref{eq:h_s}-\ref{eq:k_s}), the frequency-domain LNs $h_n(\omega)$, $l_n(\omega)$ and $k_n(\omega)$ are defined as
\begin{eqnarray}
h_n(\omega) &=& \frac{m}{a} u_n(\omega) \label{eq:h_o} \\
l_n(\omega) &=& \frac{m}{a} v_n(\omega)\label{eq:l_o}\\
k_n(\omega) &=& -1-\frac{m}{ag} \varphi_n(\omega) \label{eq:k_o}\,,
\end{eqnarray}
where $u_n(\omega)$, $v_n(\omega)$ and $\varphi_n(\omega)$ are the three components of vector $\mathbf{x}_0(\omega)=(u_n,v_n,\varphi_n)^T$. 

Since the frequency-domain LNs are complex numbers, in general {a phase difference exists} between the variation of the external periodic potential and the planet response, due to the energy dissipation within the planetary mantle. If $L_n(\omega)$ is any of the three {frequency-dependent} LNs, {the corresponding time-domain LNs are:}
\begin{equation}
L_n(t) = L_n(\omega) e^{i\omega t} = |L_n(\omega)| e^{i(\omega t - \phi)}\,,
\end{equation}
where the \textit{phase lag} $\phi$ {is}
\begin{equation}\label{eq:phaselag}
\tan\phi = -\frac{\mathrm{Im}(L_n\left(\omega\right))}{\mathrm{Re}(L_n\left(\omega\right))}\,,
\end{equation}
and $\textrm{Re}(z)$ and $\textrm{Im}(z)$ denote the real and the imaginary parts of $z$, respectively. A vanishing phase lag ($\phi=0 $) is only expected for elastic planetary models (\emph{i.e.,} for $\mathrm{Im}(L_n(\omega))=0$), {for which no dissipation occurs.} {We remark that the evaluation of the frequency-dependent Love numbers 
(\ref{eq:h_o}-\ref{eq:k_o}) does not require the application of the Post-Widder method outlined in Section \ref{sec:lntd}, since in this case no inverse transform is to be evaluated.}

Tidal dissipation is {phenomenologically} expressed in term of the \textit{quality factor} $Q$ \citep{kaula1964tidal,goldreich1966q}, which according to \emph{e.g.}
\citet{efroimsky2007tides} and \citet{clausen2015dissipation} 
is related to the phase lag $\phi$ through 
\begin{equation}\label{eq:defq}
Q(\omega) = \frac{1}{\sin \phi} = -\frac{|L_2(\omega)|}{\mathrm{Im}\, (L_2(\omega))}\,,
\end{equation}
thus implying $Q=\infty$ in the case of no dissipation. 
{
Tidal dissipation is often measured in terms of the ratio
\begin{equation}\label{eq:k2q}
\frac{|k_2|}{Q} = |k_2| \sin\phi = -\mathrm{Im}\, k_2\,.
\end{equation}
}
For terrestrial bodies, the quality factor $Q$ usually lies in a range between $10$ and $500$ \citep{goldreich1966q,murray_dermott_2000}. 
{We remark that the quality factor $Q$ is a phenomenological parameter used when the internal rheology is unknown; if LNs are computed by means of a viscoelastic model, it may be more convenient to consider the imaginary part of $k_2$, which is directly proportional to dissipation {\citep{segatz1988tidal}}}.

\section{An overview of \alma}\label{sect:alma}

{Here} we briefly outline how the solution scheme described {in previous section is} implemented {in} \alma{}, 
leaving the technical details and practical considerations to the accompanying User Manual.
\alma{} {evaluates}, for any given harmonic degree $n$, the time-domain LNs $(h_n(t), l_n(t), k_n(t))$, their time derivatives $(\dot{h}_n(t), \dot{l}_n(t), \dot{k}_n(t))$ and the frequency-domain LNs $(h_n(\omega), l_n(\omega), k_n(\omega))$, either corresponding to surface loading or to tidal boundary conditions.
{While the original version of the code was limited to time-domain LNs, the other two outputs represent new capabilities introduced by \alma{}.}
The planetary model can include, in principle, any number of layers in addition to a central core. Each of the layers can be characterized by any of the rheological laws listed in Table \ref{tab:rheologies}, while the core can also have a fluid inviscid rheology. 
{As we show in Section \ref{sect:examples} below, numerical solutions obtained with \alma{} are stable even with models including a large number of layers, providing a way to approximate rheologies whose parameters are varying continuously with radius}. 

Time-domain LNs are computed by  evaluating numerically Eqs. (\ref{eq:xtsol}) and (\ref{eq:xdtsol}), assuming a time history of the forcing that can be either a step function (Eq.~\ref{eq:timehist1}) or an elementary ramp {function} (Eq.~\ref{eq:timehist3}). In the latter case, the  duration $t_r$ of the loading phase can be configured by the user. Since Eqs. (\ref{eq:xtsol}) and (\ref{eq:xdtsol}) are singular for $t=0$, \alma{} can compute time-domain LNs only for $t>0$. In the ``elastic limit'', the LNs can be obtained 
{either by sampling them at a time $t$ that is much smaller than the characteristic relaxation times of the model, or}
by configuring the Hooke's elastic rheology for all the layers in the model. {In the second} case, the LNs will follow the same time history of the forcing. As discussed in Section \ref{sect:theory}, the sums in Eqs. (\ref{eq:xtsol}) and (\ref{eq:xdtsol}) contain oscillating terms that can lead to loss of precision due to catastrophic cancellation \citep[][]{spadaboschi2006using}. To avoid the consequent numerical degeneration of the LNs, \alma~ performs all computations in arbitrary-precision floating point arithmetic, using the Fortran FMLIB library \citep{smith1991package,smith2003using}.

When running \alma{}, the user shall configure both the number $D$ of significant digits used by the FMLIB library and the order $M$ of the Gaver sequence in Eqs. (\ref{eq:xtsol}) and (\ref{eq:xdtsol}). {As discussed by \citet{spadaboschi2006using} and \citet{spada2008alma}}, higher values of $D$ and $M$ ensure a better numerical stability and accuracy of the results, but come at the cost of rapidly increasing computation time. All the examples discussed in the next {section} have been obtained with
{parameters} $D=128$ {and} $M=8$. While these values ensure a good stability in relatively simple models, a special care shall be devoted to numerical convergence in case of models with a large number of layers and/or when computing LNs to high harmonic degrees; in that case, higher values of $D$ and $M$ may be needed to attain stable results.

{Complex-valued}
LNs are obtained by \alma{} by directly sampling Eq. (\ref{eq:xfreq1}) at the requested frequencies $\omega$, {and therefore no numerical Laplace anti-transform is performed}. {While ~for ~frequency-domain LNs the numerical instabilities associated with the Post-Widder formula are avoided, the use of high-precision arithmetic may still be appropriate, 
}
especially in case of models including a large number of layers. \alma{} does not directly compute the tidal phase lag $\phi$, the quality factor $Q$ nor the $k_2/Q$ ratio, which can be readily obtained from tabulated output values
of the real and imaginary parts of LNs through Eqs.~(\ref{eq:phaselag}-\ref{eq:k2q}).

Although \alma{} is still limited to spherically symmetric and elastically incompressible models, with respect to the version originally released by \citet{spada2008alma} now the program includes some new significant features aimed at increasing its versatility. These are: \emph{i)} the evaluation of 
frequency-dependent loading and tidal Love numbers in response to periodic forcings, \emph{ii)}, the possibility of dealing with a layered core that includes 
fluid and solid portions,
\emph{iii)} the introduction of a ramp-shaped forcing function to facilitate the implementation of loading histories varying in a 
linear piecewise manner, \emph{iv)} the implementation of the Andrade transient {viscoelastic} rheology often employed in the study of planetary deformations, \emph{v)} 
the explicit evaluation of the derivatives of the Love numbers in the time domain to facilitate the computation of geodetic variations in deglaciated areas, 
\emph{vi)} a short but exhaustive User Guide, and \emph{vii)}  
{a facilitated computation of frequency-dependent loading and tidal planetary Love numbers}, with pre-defined and easily customizable rheological profiles for some terrestrial planets and moons.

\section{Benchmarking \alma{}}\label{sect:benchmark}

{In the following} we discuss {a suite of} numerical benchmarks for Love {numbers} computed {by} \alma{}. First, we consider a uniform, incompressible, self-gravitating sphere with Maxwell rheology (the so-called ``Kelvin sphere'') and compare {tidal LNs} computed numerically by \alma{} with {well known} analytical results. 
{Then, we test numerical results from \alma{} by reproducing the viscoelastic LNs for an incompressible Earth model computed within the benchmark exercise by \citet{spada2011benchmark}.}
{Finally, we discuss the impact of the incompressibility approximation assumed in \alma{} by comparing elastic and viscoeastic LNs for a realistic Earth model with recent numerical results by \citet{Michel-and-Boy-2021}, which employ a compressible model}.

\subsection{The viscoelastic Kelvin sphere}\label{sect:benchmark_ks}

Simplified planetary models for which closed-form expressions for the LNs are available are of particular relevance here, since they allow an analytical benchmarking of the numerical solutions discussed in Section \ref{sect:theory} and provided by \alma{}. 

In what follows, we consider a spherical, homogeneous, self-gravitating model, often referred to as the ``Kelvin sphere'' \citep{thomson1863xxvii}, which can be extended to a viscoelastic rheology in a straightforward manner. 
For example, adopting {the} complex modulus $\mu(s)$ appropriate for the Maxwell rheology (see Table \ref{tab:rheologies}), for a Kelvin sphere of radius $a$, density $\rho$ and surface gravity $g$, in the Laplace domain the harmonic degree $n=2$ LNs take the form
\begin{equation}\label{eq:ksbench1}
\tilde{L}_2(s) = \frac{L_f}{\displaystyle 1+\gamma^2 \frac{s}{s+1/\tau}} \,,
\end{equation}
where $L_2$ {stands for} any of $(h_2,l_2,k_2)$, $L_f$ is the ``fluid limit'' of $\tilde{L}_2(s)$ (\emph{i.e.,} the value attained for $s\to 0$), {the Maxwell relaxation time is}
\begin{equation}
\tau=\frac{\eta}{\mu}
\end{equation} 
and
\begin{equation}\label{eq:ksbench2}
\gamma^2 = \frac{19}{2} \frac{\mu}{\rho g a}
\end{equation}
is a positive non-dimensional constant. Note that $g$ is a function of $a$ and $\rho$, since for the homogeneous sphere $g=\frac{4}{3}\pi G\rho a$, where $G$ is the {universal gravitational constant}. 

After some algebra, (\ref{eq:ksbench1}) can be cast in the form
\begin{equation}\label{eq:ksbench3}
\tilde{L}_2(s) = \frac{L_f}{1+\gamma^2}
\left( 1 + \frac{1/\tau - 1/\tau'}{s + 1/\tau'} \right)\,,
\end{equation}
where for a tidal forcing, the fluid limits for {degree}  $n=2$ are $h_f=\frac{5}{2}$, $l_f=\frac{3}{4}$ and $k_f=\frac{3}{2}$ \citep[see \emph{e.g.},][]{lambeck1988earth} and 
where we {have defined} 
\begin{equation}
\tau' = (1+\gamma^2)\tau\,.
\end{equation}

{From Eq.}~(\ref{eq:ksbench3}), the LNs in the time domain can be immediately computed analytically through an inverse {Laplace} transformation:
\begin{equation}\label{eq:ksbench4}
L_2(t) = \frac{L_f}{1+\gamma^2}
\left[ \delta(t) + H(t) \left(\frac{1}{\tau}-\frac{1}{\tau'}\right) e^{-t/\tau'} \right]\,,
\end{equation}
while for an external forcing characterized by a step-wise time-history, the LNs $L_2^{(H)}(t)$ are obtained by a 
{time convolution with the Heaviside function}:
\begin{equation}\label{eq:ksbench5a}
L_2^{(H)}(t) = L_2(t) \ast H(t)\,,
\end{equation}
{that} yields
\begin{equation}\label{eq:ksbench5}
L_2^{(H)}(t) = \frac{L_f}{1+\gamma^2} 
\left[ 1 + \gamma^2 \left( 1 - e^{-t/\tau'} \right) \right]\,, \quad t\ge 0\,,
\end{equation}
from which the time derivative of $L_2^{(H)}(t)$ is readily obtained:
\begin{equation}\label{eq:ksbench6}
\dot{L}_2^{(H)}(t) = \frac{L_f}{1+\gamma^2} 
\left( \frac{1}{\tau} - \frac{1}{\tau'} \right)
 e^{-t/\tau'} \,, \quad t> 0\,.
\end{equation}

\begin{figure}
   \begin{center}
   \includegraphics[width=10cm]{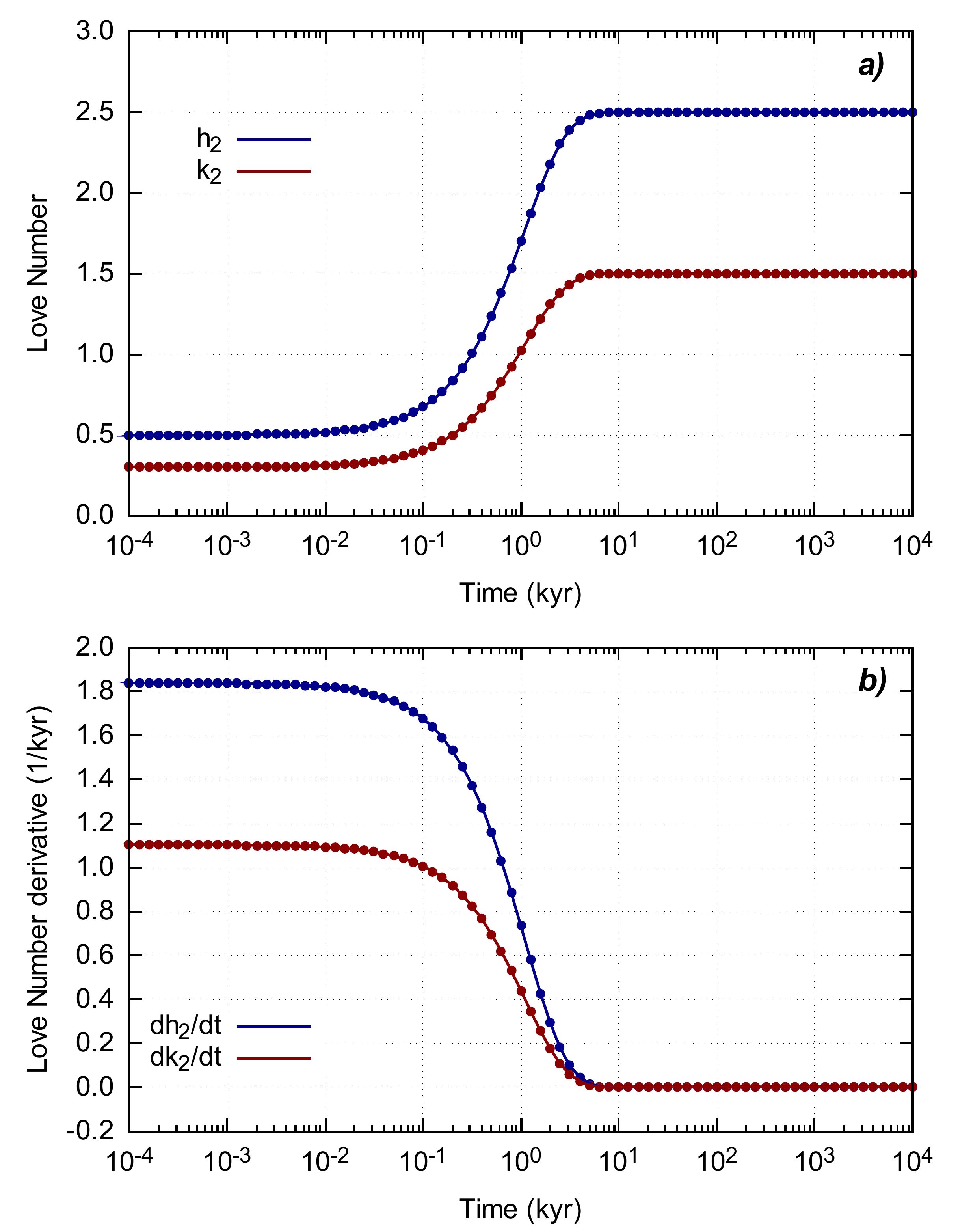}
   \end{center}
   \caption{(a) Comparison between numerical (dotted) and analytical solutions (solid) for the $h_2$ and $k_2$ tidal LNs of a Kelvin sphere with Maxwell rheology having radius $a=6371$ km, density $\rho=5.514\times 10^3$ kg$\cdot$m$^{-3}$, rigidity $\mu=1.46$ $\times$ $10^{11}$ Pa and viscosity $\eta=10^{21}$ Pa$\cdot$s. (b) The same, for the time derivatives of the LNs. Note that the time axis is logarithmic.}
   \label{fig:bench}
\end{figure}

In Figure~\ref{fig:bench}a, the \marginpar{F\ref{fig:bench}} dotted curves show the $h_2$ (blue) and the $k_2$ (red) tidal LN of harmonic degree $n=2$ obtained by a configuration of \alma~ that reproduces the Kelvin sphere (the parameters are given in the {Figure} caption).   
The LNs, shown as a function of time, are characterized by two asymptotes corresponding to the elastic and fluid limits, respectively, and by a smooth   transition in between. The solid curves, obtained by the analytical expression given by Eq.~(\ref{eq:ksbench5}), show an excellent agreement with the \alma~ numerical solutions. The same holds for the time-derivatives of these LNs, considered in Figure~\ref{fig:bench}b, where the analytical LNs (solid lines) are computed according to Eq.~(\ref{eq:ksbench6}). 

The frequency response of the Kelvin sphere for a periodic tidal potential can be obtained by setting $s=i\omega$ in Eq.~(\ref{eq:ksbench1}), which after rearranging gives:
\begin{equation}\label{eq:ksfreq2}
L_2(\omega) = \frac{L_f}{1+\gamma^2}
\left[
1+\frac{\gamma^2}{1+(\omega\tau')^2}
-i\gamma^2\frac{\omega\tau'}{1+(\omega\tau')^2}
\right]\,,
\end{equation}
which remarkably depends upon $\omega$ and $\tau$ only through the $\omega\tau$ product. Therefore, a change in the relaxation time $\tau$ shall result in a shift of the frequency response of the Kelvin sphere, leaving its shape unaltered.

Using Eq. (\ref{eq:ksfreq2}) in (\ref{eq:phaselag}), the phase lag {turns out to be}: 
\begin{equation}\label{eq:ksfreq3}
\tan\phi = \frac{\gamma^2\omega\tau}{1+\omega^2\tau\tau'}\,,
\end{equation}
where it is easy to show that for frequency 
\begin{equation}\label{eq:omega0}
\omega_0 = \frac{1}{\sqrt{\tau\tau'}}
\end{equation}
the maximum 
phase lag $\phi=\phi_{max}$ is attained, with 
\begin{equation}\label{eq:ksfreq5}
\tan\phi_{max} = \frac{\gamma^2}{2\sqrt{1+\gamma^2}}\,.
\end{equation}
By using Eq. (\ref{eq:ksfreq2}) {into} (\ref{eq:defq}), {for the Kelvin sphere the quality factor is} 
\begin{equation}\label{eq:ksfreq6}
Q_{K} (\omega) = \sqrt{
1 + \frac{1}{\gamma^4}\left( \omega\tau' + \frac{1}{\omega\tau} \right)^2}\,,
\end{equation}
which at $\omega=\omega_{0}$ attains its minimum value
\begin{equation}\label{eq:ksfreq7}
Q_{min} = 1 + \frac{2}{\gamma^2} \,.
\end{equation}

\begin{figure}
\begin{center}
\includegraphics[width=10cm]{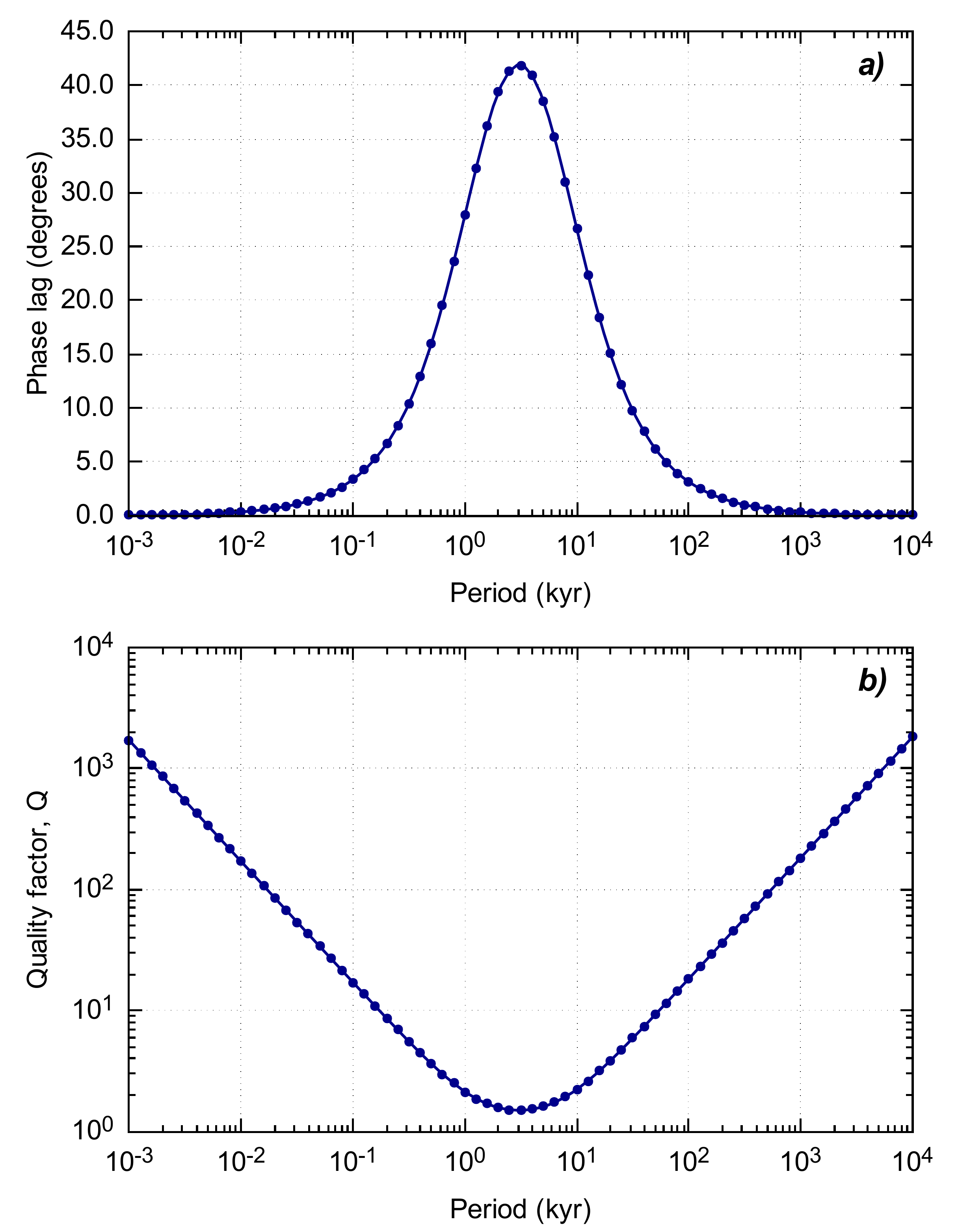}
\end{center}
\caption{Comparison between numerical (dotted) and analytical solutions (solid) for the tidal phase lag $\phi$ (a) and quality factor $Q$ (b) for the $n=2$ tidal LNs of a Kelvin sphere with Maxwell rheology, using the same parameters detailed in the caption of Figure~\ref{fig:bench}.}
\label{fig:bench2}
\end{figure}

In Figure~\ref{fig:bench2}a, \marginpar{F\ref{fig:bench2}} the dotted curve shows the phase lag $\phi$ as a function of the tidal period $T=2\pi/\omega$, obtained by the same configuration of \alma~ described in the caption of Figure~\ref{fig:bench}. The solid line corresponds to the analytical expression of $\phi(T)$ which can be obtained from Eq.~(\ref{eq:ksfreq3}), showing once again an excellent agreement with the numerical results (dotted). Figure~\ref{fig:bench2}b compares numerical results obtained from \alma~ for $Q$ with the analytical expression for $Q_K(T)$ obtained from (\ref{eq:ksfreq6}). By using in Eq.~(\ref{eq:omega0}) the numerical values of $\rho$, $a$ and $\mu$ assumed in Figures~\ref{fig:bench} and \ref{fig:bench2}, the period $T_0=2\pi/\omega_0$ is found to scale with viscosity $\eta$ as
\begin{equation}\label{eq:ksfreq8}
T_0 = (3.06\,\mathrm{kyr}) \left( \frac{\eta}{10^{21}\, \mathrm{Pa\cdot s}} \right)\,,
\end{equation}
so that for $\eta=10^{21}\,\mathrm{Pa\cdot s}$, representative of the Earth's mantle bulk viscosity \citep[see \emph{e.g.,}][]{mitrovica1996haskell,turcotte2014geodynamics}, the maximum phase lag $\phi_{max}\simeq 41.9^\circ$ and the minimum quality factor $Q_{min} \simeq 1.5 $ are attained for $T_0 \simeq 3\,\mathrm{kyr}$, consistent with the results shown in Figure~\ref{fig:bench2}.

\subsection{Community-agreed Love numbers for an incompressible Earth model}\label{sect:benchmark_spada2011}

{
Due to the relevance of viscoelastic Love numbers in a wide range of applications in Earth science, several numerical approaches for their evaluation have been independently developed and proposed in literature. This ignited the interest on benchmark exercises, in which a set of agreed numerical results can be obtained and different approaches and methods can be cross-validated. Here we consider a benchmark effort that has taken place in the framework of the Glacio-Isostatic Adjustment community \citep{spada2011benchmark}, in which a set of reference viscoelastic Love numbers for an incompressible, spherically symmetric Earth model has been derived through different numerical approaches, including viscoelastic normal modes, spectral-finite elements and finite elements. This allows us to validate our numerical results by implementing in \alma{} the M3-L70-V01 Earth model described in Table $3$ of \citet{spada2011benchmark}, which includes a fluid inviscid core, three mantle layers with Maxwell viscoelastic rheology and an elastic lithosphere, and comparing the set of LNs from \alma{} with reference results from the benchmark exercise.
}

\begin{figure}
   \begin{center}
   \includegraphics[width=15cm]{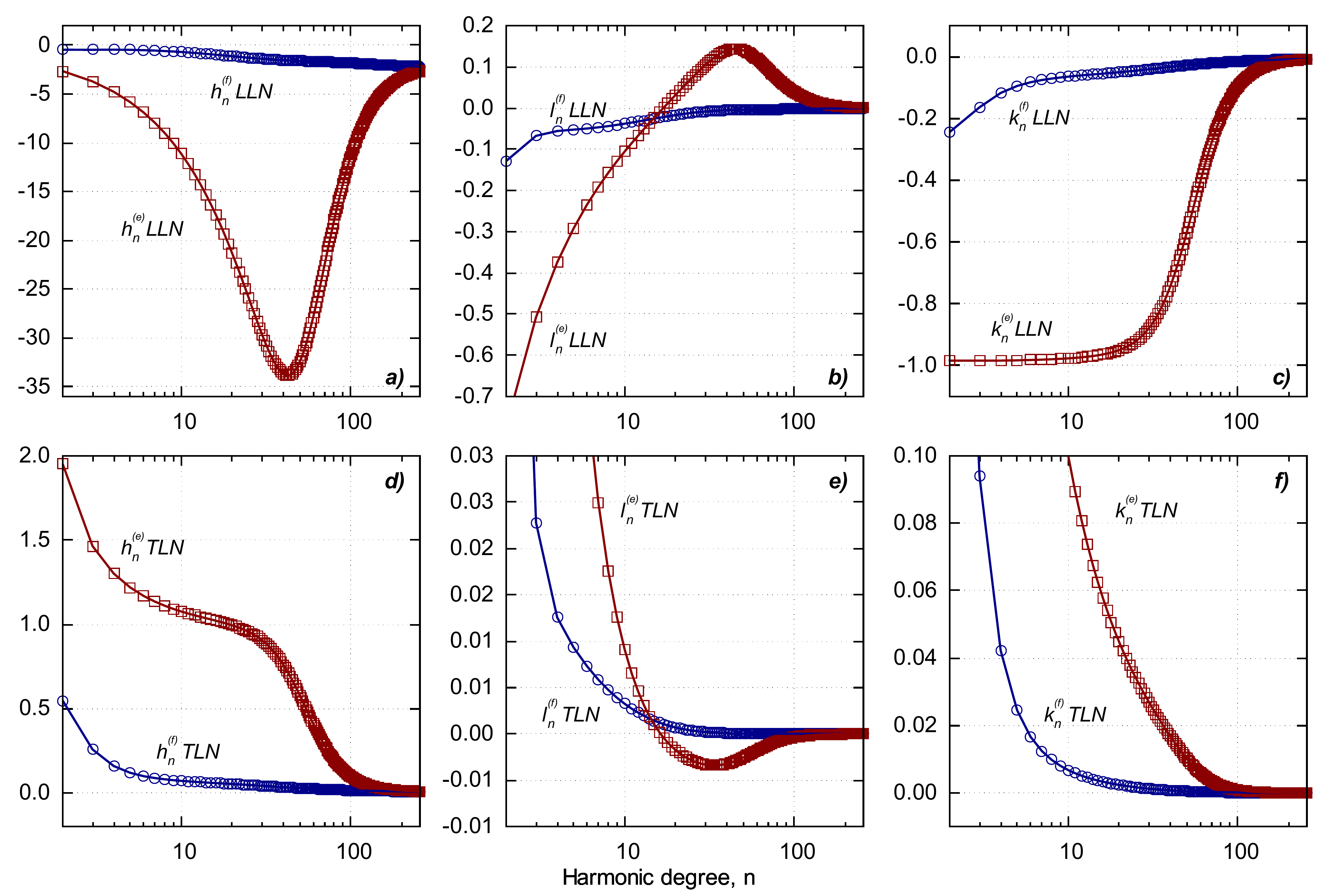}
   \end{center}
   \caption{
{
Elastic (red) and fluid (blue) Love numbers as a function of the harmonic degree for the Earth model M3-L70-V01 defined in \citet{spada2011benchmark}. Top (a-c) and bottom frames (d-f) show Love numbers for loading and tidal forcing, respectively. Symbols show numerical results obtained with \alma{} while solid lines represent reference results from the benchmark exercise by \citet{spada2011benchmark}. 
}   
   }
   \label{fig:bench2011-1}
\end{figure}

{
Figure \ref{fig:bench2011-1}\marginpar{F\ref{fig:bench2011-1}} shows elastic ($h_n^{(e)}$, $l_n^{(e)}$, $k_n^{(e)}$) and fluid LNs ($h_n^{(f)}$, $l_n^{(f)}$, $k_n^{(f)}$), both for the loading and tidal cases, computed by \alma{} for the M3-L70-V01 Earth model in the range of harmonic degrees $2\le n\le 250$. 
The elastic and fluid limits have been simulated in \alma{} by sampling the time-dependent LNs at $t_{e}=10^{-5}$ kyrs and $t_{f}=10^{10}$ kyrs, respectively. Reference results from \citet{spada2011benchmark}, represented by solid lines in Figure \ref{fig:bench2011-1}, are practically indistinguishable from results obtained with \alma{} over the whole range of harmonic degrees, demonstrating the reliability of the numerical approach employed in \alma{}.
}

\begin{figure}
   \begin{center}
   \includegraphics[width=15cm]{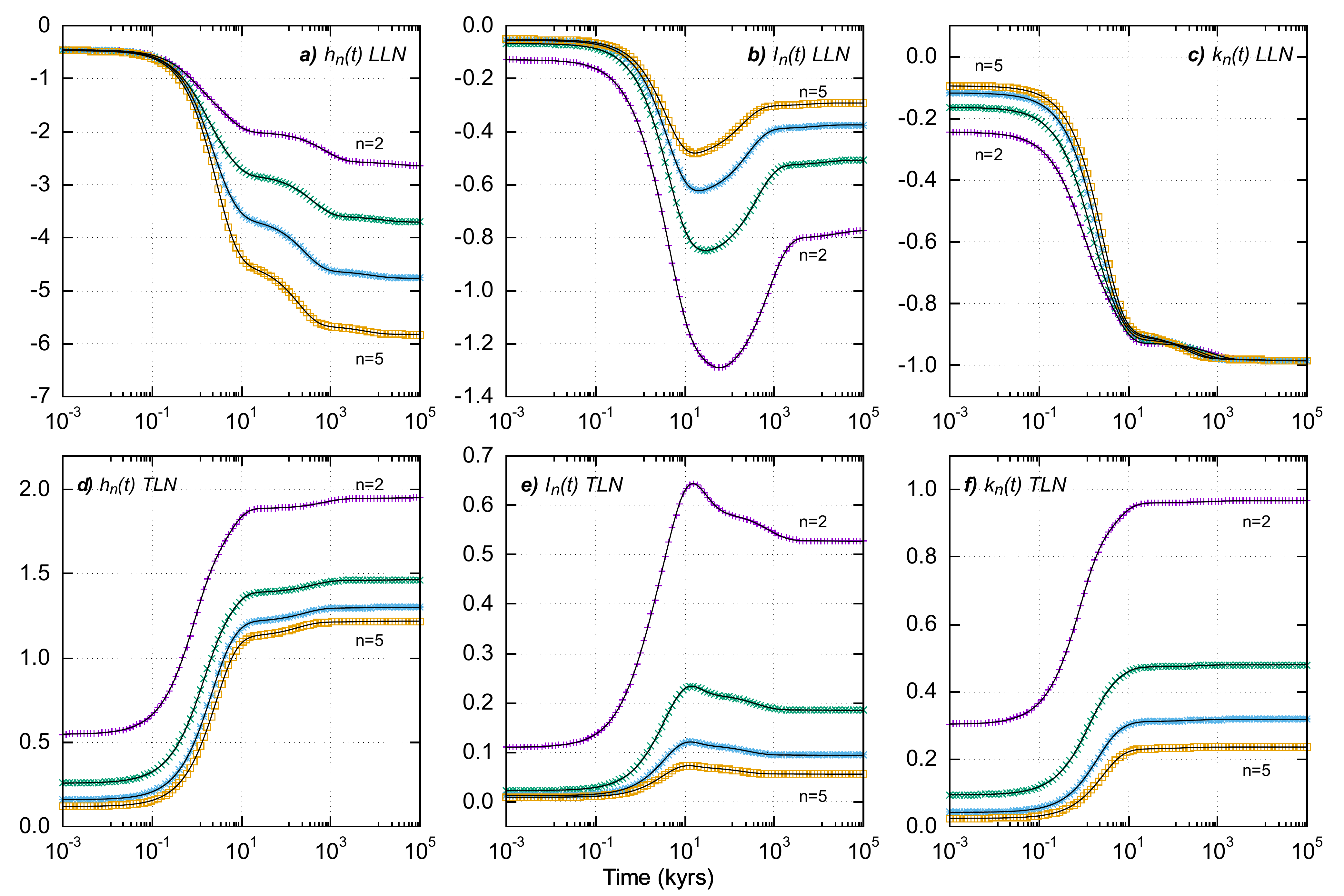}
   \end{center}
   \caption{
{
Time-dependent viscoelastic Love numbers for the M3-L70-V01 Earth model at long spatial wavelength (harmonic degrees $2\le n\le 5$).
Top panels (frames a-c) and bottom ones (d-f) show Love numbers for a loading and tidal forcing, respectively. The time history of the load is an Heaviside unit step function.
Symbols show numerical results obtained with \alma{} while solid lines represent reference results from the benchmark exercise by \citet{spada2011benchmark}.
}   
   }
   \label{fig:bench2011-2}
\end{figure}

{
Figure \ref{fig:bench2011-2}\marginpar{F\ref{fig:bench2011-2}} shows time-dependent LNs $h_n(t)$, $l_n(t)$ and $k_n(t)$, for both the loading and tidal cases, computed by \alma{} for harmonic degrees $2\le n \le 5$ and for $t$ between $10^{-3}$ and $10^5$ kyrs, a time range that encompasses the complete transition between the elastic and fluid limits. Also in this case, numerical results obtained by \alma{} (shown by symbols) are coincident with the reference LNs from \citet{spada2011benchmark}, represented by solid lines.
}

\subsection{Viscoleastic Love numbers for a PREM-layered Earth model}\label{sect:benchmark_prem}

{In this last benchmark, we compare numerical results} from \alma~ with reference viscoelastic LNs for a realistic Earth model which accounts for an elastically compressible rheology, in order to assess its importance when modeling the tidal and loading response of a large planetary body. In the context of
Earth rotation, the role of compressibility has been addressed by \citet{Vermeersen1996compressible}; the reader is also referred to \citet{sabadini2016global} for a broader presentation of the problem
and to \citet{renaud2018increased} for a discussion of the effects of compressibility in the realm of planetary modelling.

{Here we focus on numerical results recently obtained by \citet{Michel-and-Boy-2021}, who employed Fourier techniques to compute frequency-dependent viscoelastic LNs for periodic forcings both of loading and tidal types. They have adopted an Earth model with the elastic structure of PREM \citep[Preliminary Reference Earth Model, ][]{Dziewonski1981Preliminary} and a fully liquid core, and replaced the outer oceanic layer with a solid crust layer, adjusting crustal density in such a way to keep the total Earth mass unchanged. Following \citet{Michel-and-Boy-2021}, we have built a discretised realization of PREM suitable for \alma~ with {a fluid core and} $28$ homogeneous {mantle} layers, which has been used to obtain the numerical results discussed below.}

\begin{figure}
\begin{center}
\includegraphics[width=15
cm]{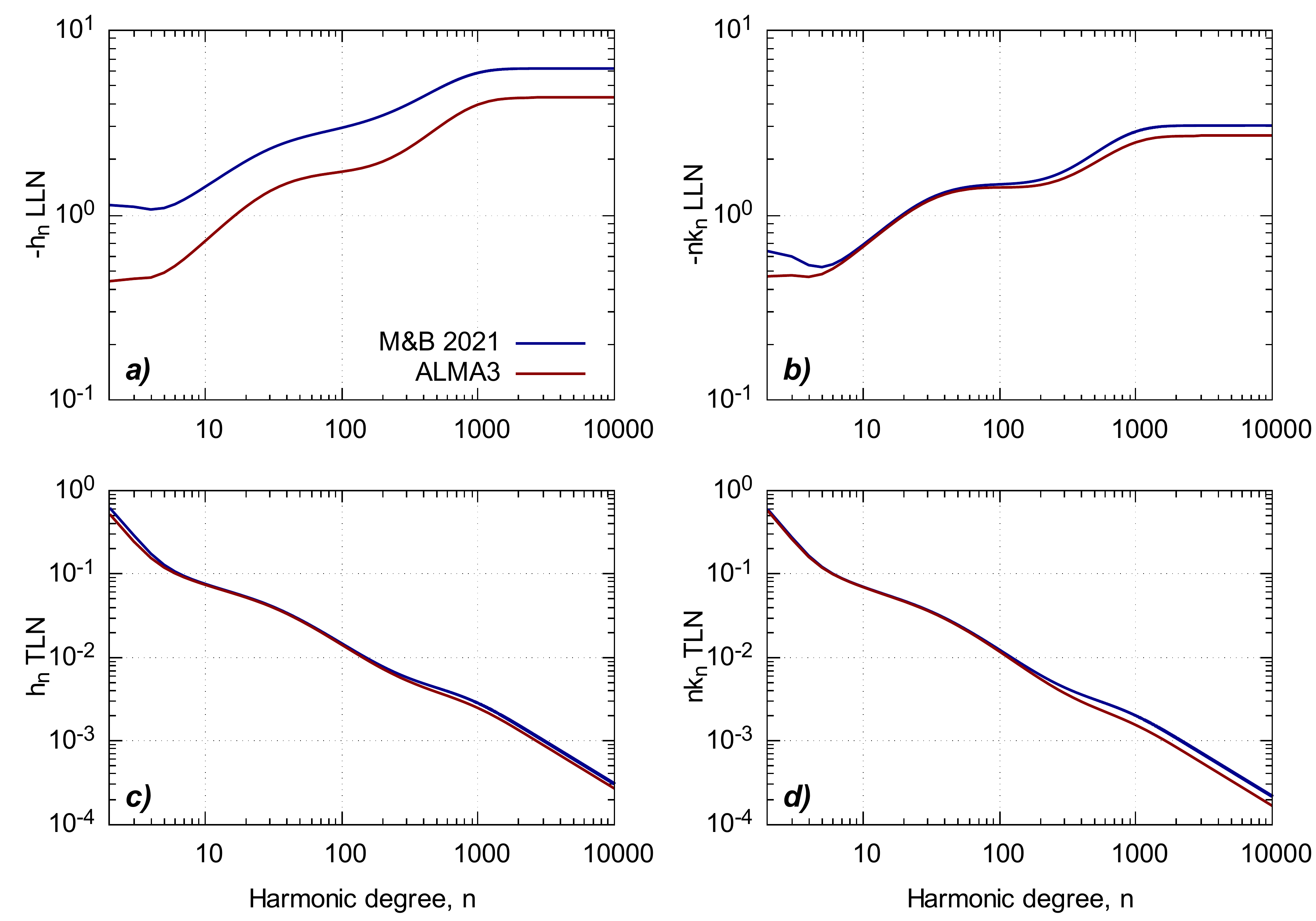}
\end{center}
\caption{
{
Comparison between elastic Love numbers $h_n$ (left panels) and $k_n$ (right) obtained by \citet{Michel-and-Boy-2021} with numerical results from \alma. In both cases, the Earth model has the elastic structure of PREM {in the crust and in the mantle}, while the core is modeled as an uniform, inviscid fluid. Top and bottom panels show loading and tidal Love numbers, respectively.}
}
\label{fig:bench3}
\end{figure}

{
Figure \ref{fig:bench3} \marginpar{F\ref{fig:bench3}} compares elastic Love numbers obtained by \citet{Michel-and-Boy-2021} in the range of harmonic degrees between $n=2$ and $n=10,000$ with those computed with \alma~. The largest difference between the two sets of LNs can be seen for $h_n$ in the loading case (Figure \ref{fig:bench3}a), where the assumption of incompressibility leads to a significant underestimation of deformation across the whole range of harmonic degrees. Incompressible elasticity leads to an underestimation also of the $k_n$ loading LN (Figure \ref{fig:bench3}b), although the differences are much smaller and limited to the lowest harmonic degrees. Conversely, for the tidal response (Figures~\ref{fig:bench3}c and \ref{fig:bench3}d) the two sets of LNs turn out to be almost overlapping, suggesting a minor impact of elastic compressibility on tidal deformations.}

\begin{figure}
\begin{center}
\includegraphics[width=15cm]{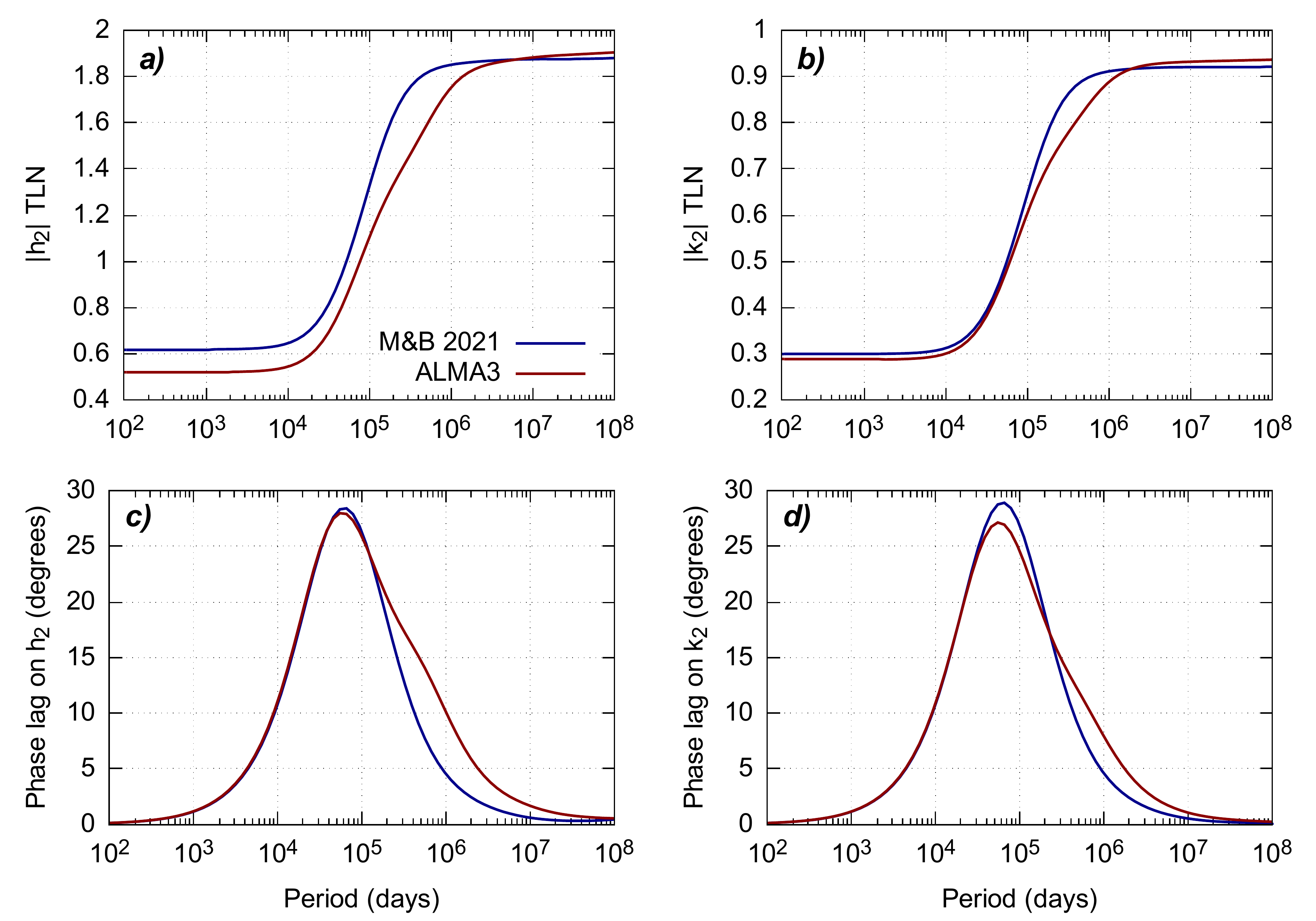}
\end{center}
\caption{
{
Comparison between the modulus of tidal Love numbers $|h_2|$ and $|k_2|$ (top panels) and corresponding phase lags (bottom panels) obtained by \citet{Michel-and-Boy-2021} for a periodic forcing with numerical results from  \alma~. The Earth model has the elastic structure of PREM and a Maxwell rheology with viscosity $\eta=10^{21}$ Pa$\cdot$s is assumed in the mantle.}
}
\label{fig:bench4}
\end{figure}

{
In Figure~\ref{fig:bench4} \marginpar{F\ref{fig:bench4}} we consider a periodic load and compare viscoelastic tidal LNs $h_2$ and $k_2$ computed with \alma~ with corresponding results from \citet{Michel-and-Boy-2021}. Consistently with the elastic case, we see that the incompressibility approximation used in \alma~ generally results in smaller modeled deformations across the whole range of forcing periods. The largest differences are found on $|h_2|$ (Figure \ref{fig:bench4}a) and reach the $\sim 20\%$ level in the range of periods between $10^5$ and $10^6$ days, while on $|k_2|$ (\ref{fig:bench4}b) the differences are much smaller, reaching the $\sim 10\%$ level in the same range of periods. Similarly, for the phase lags (Figures \ref{fig:bench4}c and \ref{fig:bench4}d) we find a larger difference for $h_2$ than for $k_2$, with the phase lag being remarkably insensitive 
{to compressibility}
up to forcing periods {of the order of $10^4$-$10^5$ days}.
}

\section{Examples of \alma{} applications}\label{sect:examples}

{In this Section}
we consider four applications showing the potential of \alma~ in different contexts.  
{First, we will discuss the $k_2$ tidal Love number of Venus}, based upon a realistic layering for the interior of this planet. 
Second, we shall evaluate the tidal LNs for a simple model of the Saturn's moon Enceladus, in order to show how an internal fluid layer can be simulated as a low-viscosity Newtonian fluid rheology {and how a depth-dependent viscosity in a conductive shell may be approximated using a sequence of thin homogeneous layers}. Third, we will evaluate a set of loading LNs suitable for describing the transient response of the Earth to the melting of large continental ice sheets. As a last example, we will demonstrate how \alma~ can simulate the tidal dissipation on the Moon using two recent interior models based on seismological data. While these numerical experiments are put in the context of state-of-the-art planetary interior modeling, we remark that they are aimed only at illustrating the modeling capabilities of \alma~.

\subsection{Tidal deformation of Venus}\label{sec:venus}

{The planet Venus is often referred to as ``Earth's twin planet'', since its size and density differ only by $\sim 5\%$ from those of the Earth. These similarities lead to the expectation that the chemical composition of the Earth and Venus may be similar, with an iron-rich core, a magnesium 
silicate mantle and a silicate crust \citep{kovach1965interiors,lewis1972metal,Anderson1980Tectonics}. Despite these similarities, there is a lack of constraints on the internal structure of Venus. Therefore, its density and rigidity profiles are often assumed to be a re--scaled version of the Preliminary Reference Earth Model (PREM) of \citet{Dziewonski1981Preliminary}, accounting for the difference in the planet's radius and mass, as in~\citet{aitta2012venus}. One of the main observational constraints on the planet's interior, along its mass and moment of inertia, is its $k_2$ tidal LN. The current observational estimate of $k_2$ for Venus is $0.295 \pm 0.066$ ($2\times \mathrm{formal}\ \sigma$), {and it has been} inferred from Magellan and Pioneer Venus orbiter spacecraft data \citep{konopliv1996venusian}. However, due to uncertainties on $k_2$, it is not possible to discriminate between a liquid and a solid core~\citep{dumoulin2017tidal}}.

\begin{figure}
\begin{center}
\includegraphics[scale=0.18]{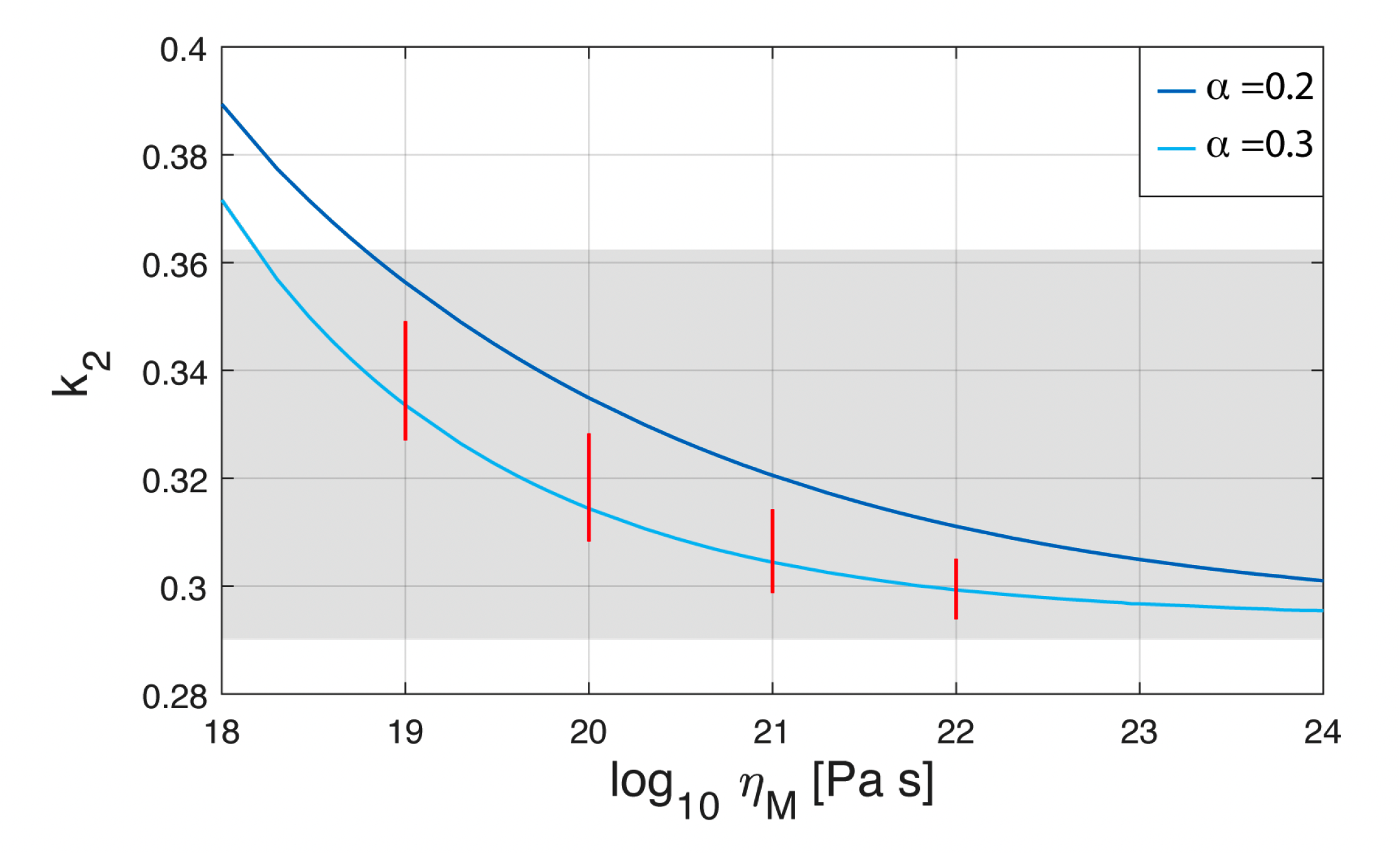}
\end{center}
\caption{
{
{
Tidal Love Number $k_2$ as a function of the mantle viscosity $\eta_M$ for the internal model $T_5^{hot}$ of \cite{dumoulin2017tidal}. The two curves correspond to numerical results from \alma{} assuming Andrade creep parameters $\alpha=0.2$ and $\alpha=0.3$, respectively. Red vertical segments represent the range of the estimates obtained by~\cite{dumoulin2017tidal}, while the grey shaded area represents the most recent observed value of $k_2$ and its $2\sigma$ uncertainty according to~\citet{konopliv1996venusian}. 
}
}
}
\label{Fig:Venus_k2_Q}
\end{figure}

{
Here we use \alma~ to reproduce results obtained by means of the Venus model referred to as $T_5^{hot}$ by \citet{dumoulin2017tidal}, based on the ``hot temperature profile'' from \citet{Armann2012Simulating}, having a composition and hydrostatic pressure from the PREM model of \citet{Dziewonski1981Preliminary}. The viscosity $\eta$ of the mantle of Venus is fixed and homogeneous; the crust is elastic ($\eta \rightarrow \infty$), the core is assumed to be inviscid ($\eta = 0$) and the rheology of the mantle follows Andrade's law (see Table \ref{tab:rheologies}).
The parameters of the $T_5^{hot}$ {model} have been volume-averaged into the core, the lower mantle, the upper mantle and the crust. 
The calculation of $k_2$ is {performed at the} tidal period of $58.4$ days \citep{cottereau2011various}. In the work of \citet{dumoulin2017tidal}, $k_2$ is computed by integrating the radial functions associated with the gravitational potential, as defined by \citet{takeuchi1972seismic},
hence the simplified formulation of \citet{Saito1974Some} relying on the radial function is employed. The method is derived from the classical theory of elastic body deformation and the energy density integrals commonly used in the seismological community. One of the main differences between their computation and the results presented here is the assumption about compressibility, since \citet{dumoulin2017tidal} use a compressible planetary model, while in \alma~ an incompressible rheology is always assumed. 
{
In \marginpar{F\ref{Fig:Venus_k2_Q}} Figure~\ref{Fig:Venus_k2_Q}, the two curves show the $k_2$ tidal LN corresponding to Andrade creep parameters $\alpha=0.2$ and $\alpha=0.3$ as a function of mantle viscosity for the tidal period of $58.4$ days. 
}
Each of the vertical red segments {corresponds to} the interval of $k_2$ values obtained by \citet{dumoulin2017tidal} for discrete mantle viscosity values $\eta = 10^{19}$, $10^{20}$, $10^{21}$ and $10^{22}$ Pa$\cdot$s, {respectively, and} for a range of the Andrade creep parameter $\alpha$ in the interval  between $0.2$ and $0.3$. The grey shaded area illustrates the most recent observed value of $k_2$ according to~\citet{konopliv1996venusian} to an uncertainty of $2 \times\ \mathrm{formal}\ \sigma$. 
Figure~\ref{Fig:Venus_k2_Q} shows that the $k_2$ values obtained with \alma~ for the $T_5^{hot}$ Venus model fit well with the lower boundary of the compared study for each of the discrete mantle viscosity values {if an Andrade creep parameter $\alpha=0.3$ is assumed, while for $\alpha=0.2$ the modeled $k_2$ slightly exceeds the upper boundary of \citet{dumoulin2017tidal}}.
}

\subsection{The tidal response of Enceladus}

The scientific interest on Enceladus has gained considerable momentum after the 2005 Cassini flybys, which confirmed the icy nature of its surface and evidenced the existence of water-rich plumes emerging from the southern polar regions \citep{porco2006cassini,ivins2020linear}. These hint to the existence of a subsurface ocean, heated by tidal dissipation in the core, where physical conditions allowing life could be possible, in principle~\citep[for a review, see][]{hemingway2018interior}. The interior structure of Enceladus has been thoroughly investigated in literature on the basis of observations of its gravity field \citep{iess2014gravity}, tidal deformation and physical librations \citep[see, \textit{e.g.}][]{cadek2016enceladus}, setting constraints on the possible structure of the ice shell and of the underlying liquid ocean \citep{roberts2008tidal}, and on the composition of its core \citep{roberts2015fluffy}.
{Lateral variations in the crustal thickness of Enceladus have been inferred in studies about the isostatic response of the satellite using gravity and topography data as constraints~\citep[see][]{cadek2016enceladus,beuthe2016enceladus,CADEK2019476} and in works dealing with the computation of deformation and dissipation \citep[see][]{souvcek2016effect,souvcek2019tidal,beuthe2018enceladus,beuthe2019enceladus}. Indeed, from all the  above studies, it clearly emerges that a full insight into the tidal dynamics of Enceladus could be only gained adopting 3D models of its internal structure. }

While a thorough investigation of the signature of the interior structure of Enceladus on its tidal response is far beyond the scope of this work, here we set up a simple {spherically symmetric} model with the purpose of illustrating how the LNs for a planetary body including a fluid internal layer like Enceladus can be computed with \alma, {and how a radially-dependent viscosity structure can be approximated with homogeneous layers}. 
{We define a spherically symmetric model including an homogeneous inner solid core of radius $c=192$ km \citep{hemingway2018interior}, surrounded by a liquid water layer and an outer icy shell, and investigate the sensitivity of the tidal LNs to the thickness of the ice layer, along the lines of \citet{roberts2008tidal} and \citet{beuthe2018enceladus}. 
In our setup, the core is modeled as a homogeneous elastic body with rigidity $\mu_c=4\times 10^{10}$ Pa and whose density is adjusted to ensure that, when varying the thickness of the ice shell, the average bulk density of the model is kept constant at $\rho_b = 1610$ kg$\cdot$m${}^{-3}$. Since in \alma~ a fluid inviscid rheology can be prescribed only for the core, we approximate the ocean layer {as a low viscosity Newtonian fluid ($\eta_w=10^4$ Pa$\cdot$s)}. The ice shell is modeled as a conductive Maxwell body whose viscosity profile depends on the temperature $T$ according to the Arrhenius law}:
\begin{equation}\label{eq:arrhenius}
\eta(T) = \eta_m \exp\left[ \frac{E_a}{R_g T_m} \left(\frac{T_m}{T} - 1\right)\right]\,,
\end{equation}
{where $E_a$ is the activation energy, $R_g$ is the gas constant, $T_m$ is the temperature at the base if the ice shell and $\eta_m$ is the ice viscosity at $T=T_m$. Following \citet{beuthe2018enceladus}, we use $E_a=59.4\, \mathrm{J/(mol\cdot K)}$, $\eta_m=10^{13}$ Pa$\cdot$s and $T_m=273\, \mathrm{K}$, and assume that the temperature  inside the ice shell varies with radius $r$ according to}
\begin{equation}\label{eq:temp_vs_r}
T(r) = T_m^{\frac{r-a}{r_b-a}} T_s^{\frac{r_b-r}{r_b-a}}\,,
\end{equation}
{where $r_b$ is the bottom radius of the ice shell and $T_s=59\,\mathrm{K}$ is the average surface temperature. Since in \alma{} the rheological parameters must be consant inside each layer, we discretize the radial viscosity profile given by Eq. (\ref{eq:arrhenius}) using a onion-like structure of homogeneous spherical shells. To assess the sensitivity of results to the choice of discretization resolution, we perform three numerical experiments in which the thickness of ice layers is set to $0.25$, $0.5$ and $1$ km. The ice and water densities are set to $\rho_i=930$  kg$\cdot$m${}^{-3}$ and $\rho_w=1020$ kg$\cdot$m${}^{-3}$, respectively, while the ice rigidity is set to $\mu_i=3.5\times 10^9$ Pa, a value consistent with evidence from tidal flexure of marine ice \citep{vaughan1995flexure} and laboratory experiments \citep{cole1995cycling}. }

\begin{figure}
\begin{center}
\includegraphics[width=10cm]{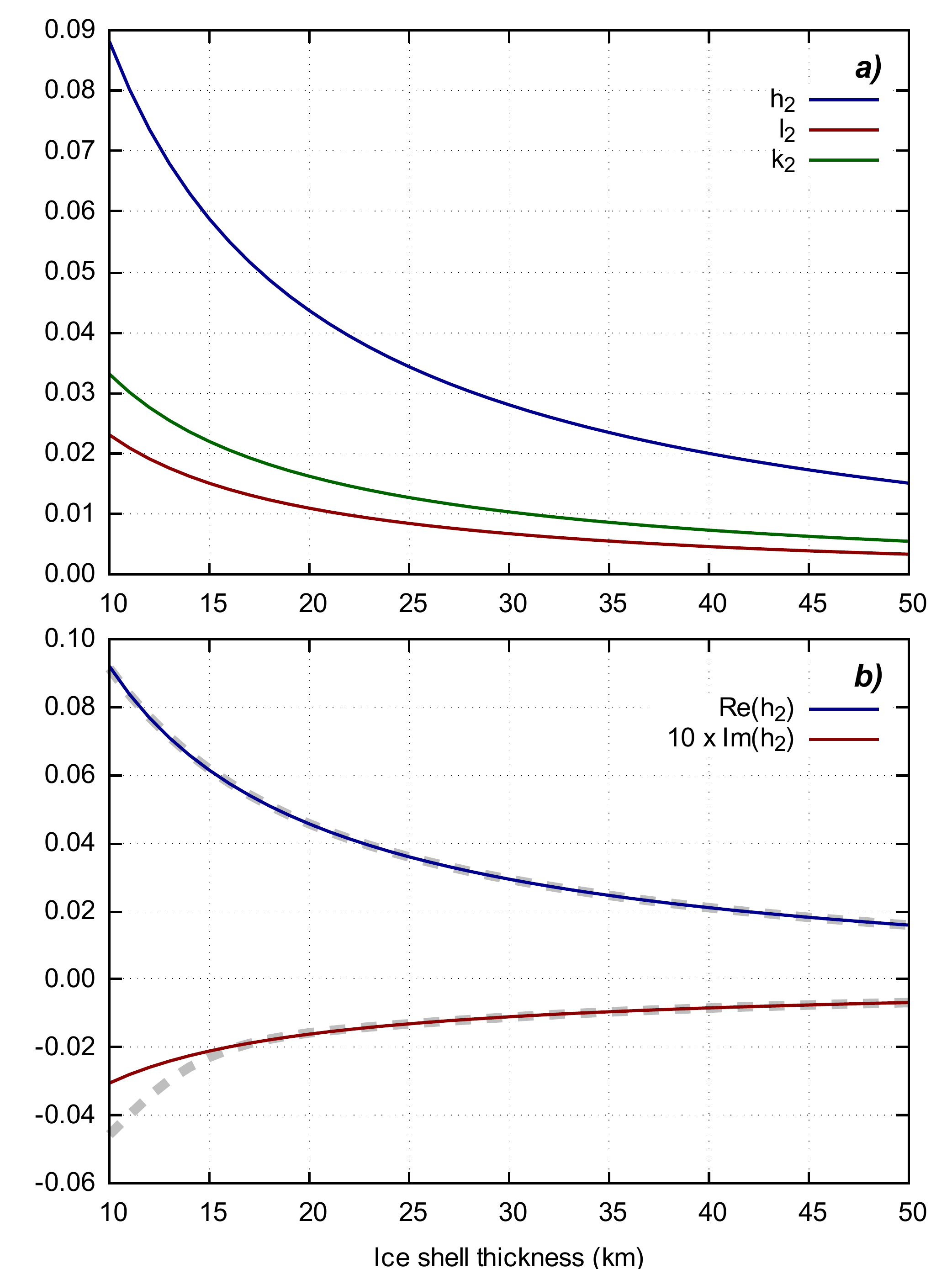}
\end{center}
\caption{
{
Elastic harmonic degree $2$ tidal Love numbers for Enceladus (a) as a function of the thickness of the ice shell. In (b), real and imaginary parts of the viscoelastic tidal Love number $h_2$ for a forcing period of $1.73$ days are shown. Solid lines and dashed lines correspond to discretization steps for the ice shell of $0.50$ and $1.00$ km, respectively. Please note that $\mathrm{Im}(k_2)$ has been multiplied by a factor of $10$ to improve readability.}
}
\label{fig:enceladus-1}
\end{figure}

{
Figure~\ref{fig:enceladus-1}a shows the elastic tidal LNs $h_2$, $l_2$ and $k_2$ for the Enceladus model discussed above as a function of the thickness of the ice shell. The elastic tidal response is strongly dependent on the ice thickness, with the $h_2$ LNs decreasing from $\sim 0.090$ for a $10$~km-thick shell to $\sim 0.015$ for a $50$~km-thick shell. It is of interest to compare these results with elastic LNs obtained by \citet{beuthe2018enceladus} in the uniform-shell approximation. It turns out that the $h_2$ LN shown in Figure~\ref{fig:enceladus-1}a is slightly smaller than corresponding results from \citet{beuthe2018enceladus}, with relative differences between the $5$-$10\%$ level, consistently with their estimate of the effect of incompressibility. Figure~\ref{fig:enceladus-1}b shows the real and imaginary parts of the $h_2$ tidal LN as a function of the thickness of the ice layer for a periodic load of period $T=1.37$ days, which corresponds to the shortest librational oscillation of Enceladus \citep{rambaux2010enceladus}. As discussed above, for this numerical experiment we implemented in \alma{} a radially-variable viscosity profile by discretizing Eq. (\ref{eq:arrhenius}) into a series of uniform layers. Solid and dashed lines in Figure~\ref{fig:enceladus-1}b show results obtained with a discretization step of $0.5$ km and $1.0$ km, respectively; we verified that with a step of $0.25$ km the results are virtually identical to those obtained with a step of $0.5$ km. 
The effect of the discretization is evident only on the imaginary part of $h_2$, where a coarse layer size of $1$ km leads to a significant overestimation of $\mathrm{Im}(k_2)$ if the ice shell is thinner than $\sim 15$ km. 
By a visual comparison of the results of Figure~\ref{fig:enceladus-1}b with Figure 4 of \citet{beuthe2018enceladus}, we can see that the imaginary part of $h_2$ is well reproduced, while the real part is underestimated by the same level we found for the elastic LNs; this difference is likely to be attributed to the incompressibility approximation adopted in \alma{}. 
}

\subsection{Loading Love numbers for transient rheologies in the Earth's mantle}

Loading Love numbers are key components in models of the response of the Earth to the spatio-temporal variation of surface loads, including the ongoing deformation due to the melting of the late Pleistocene ice complexes \citep[see \emph{e.g.,}][]{peltier2008rheological,purcell2016assessment}, the present-day and future response to climate-driven melting of ice sheets and glaciers \citep[][]{bamber2010sea,Slangen_2012}, and deformations induced by the variation of hydrological loads \citep{bevis2016on,silverii2016aquifers}. Evidence from Global Navigation Satellite System  measurements of the time-dependent surface deformation point to a possible transient nature of the mantle in response to the regional-scale melting of ice sheets and to large earthquakes \citep[see, \textit{e.g.}, ][]{pollitz2003transient,pollitz2005transient,nield2014rapid,qiu2018transient}. Here,
it is therefore of interest to present the outcomes of some numerical experiments in which \alma~ is configured to compute the time-dependent $h$ loading Love Number assuming a transient rheology in the mantle. Numerical estimates of $h_n(t)$ and of its time derivative $\dot{h}_n(t)$ would be needed, for instance, to model the response to the thickness variation of a disc-shaped surface load, as discussed by \citet{bevis2016on}.

\begin{figure}
\begin{center}
\includegraphics[width=10cm]{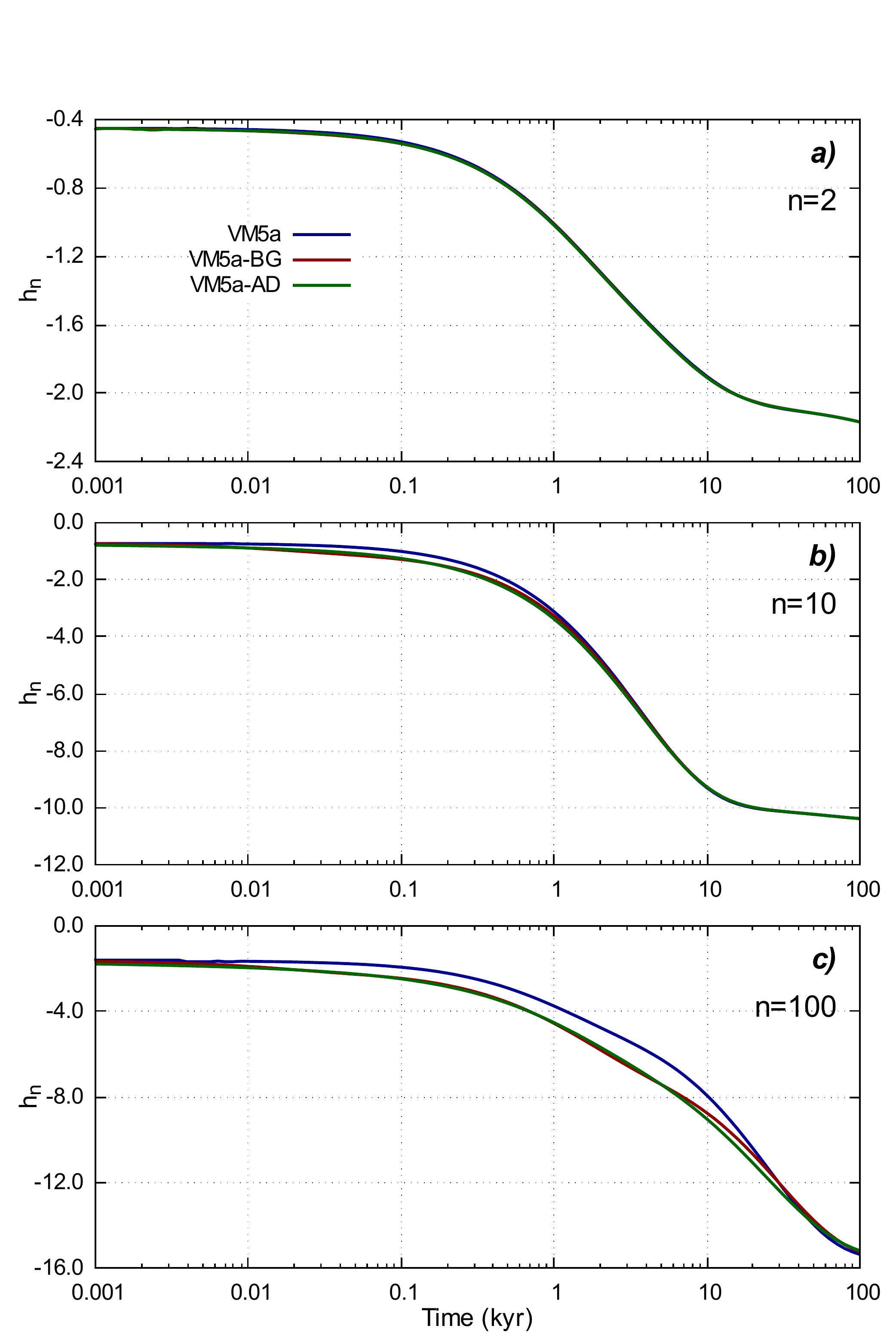}
\end{center}
\caption{Loading Love number $h_n(t)$ for $n=2$ (frame a), $n=10$ (b) and $n=100$ (c), obtained with the VM5a viscosity model by \citet{peltier2008rheological} and with two variants that assume Burgers (VM5a-BG) or Andrade (VM5a-AD) rheologies in the upper mantle layers.}
\label{fig:earth-loading-1}
\end{figure}

\begin{figure}
\begin{center}
\includegraphics[width=10cm]{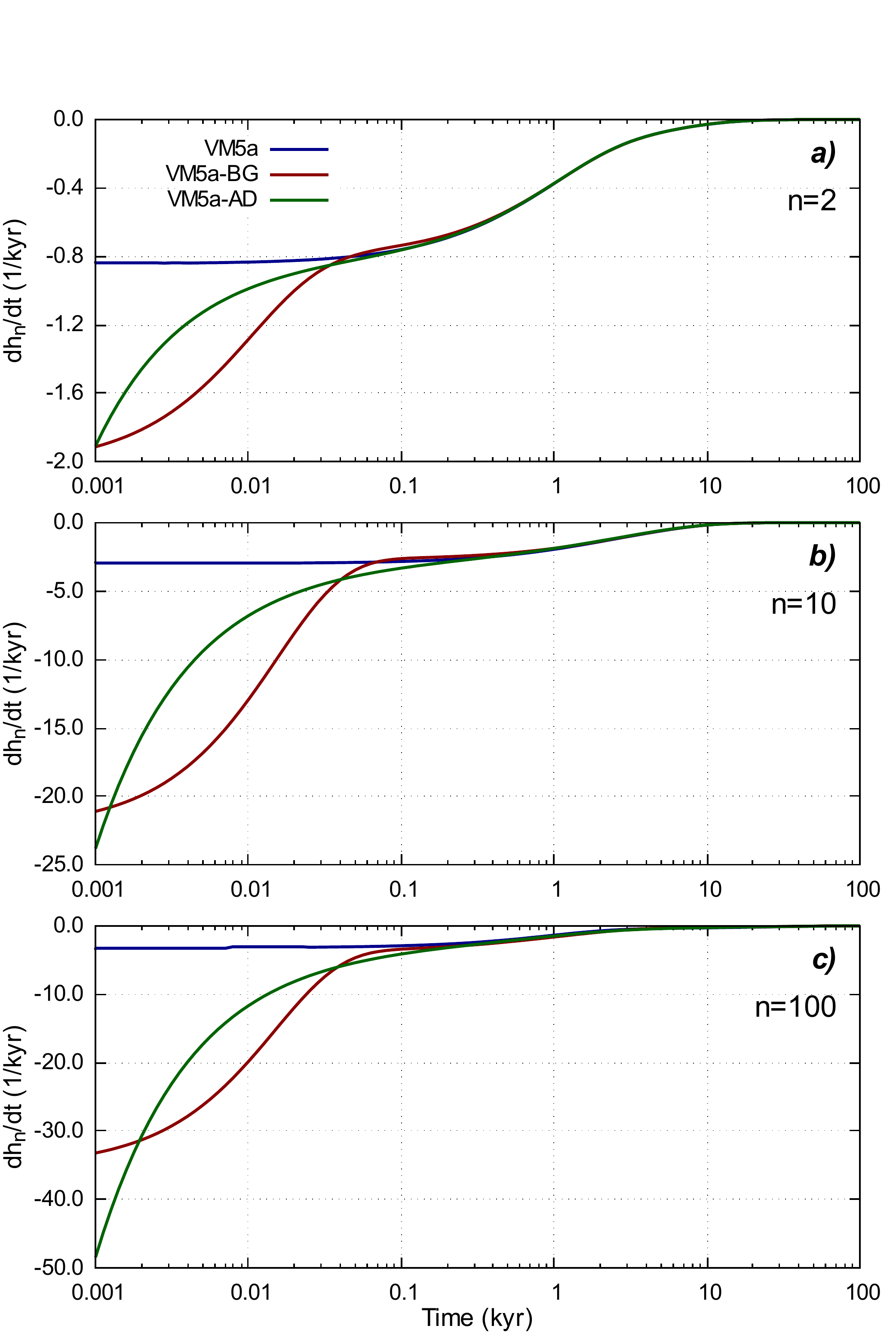}
\end{center}
\caption{Time-derivative of the loading Love number $\dot{h}_n(t)$ for {harmonic degrees} $n=2, 10$ and $100$, {adopting} the rheological models described in the caption of Figure~\ref{fig:earth-loading-1}.}
\label{fig:earth-loading-2}
\end{figure}

In Figure~\ref{fig:earth-loading-1} we \marginpar{F\ref{fig:earth-loading-1}} show the time evolution of the $h_n(t)$ loading LN for $n=2, 10$ and $100$, comparing the response obtained assuming the VM5a viscosity model of \citet{peltier2008rheological}, which is fully based on a Maxwell rheology, with those expected if VM5a is modified introducing a transient rheology in the upper mantle layers. An Heaviside time history for the load is adopted throughout. In model VM5a-BG we assumed a Burgers bi-viscous rheological law in the upper mantle, with $\mu_2=\mu_1$ and $\eta_2/\eta_1=0.1$ (see Table \ref{tab:rheologies}), while in model VM5a-AD an Andrade rheology \citep{doi:10.1080/095008396181082} with creep parameter $\alpha=0.3$ has been assumed for the upper mantle. For $n=2$ (Figure \ref{fig:earth-loading-1}a) the responses obtained with the three models almost overlap. Indeed, for long wavelengths (by Jean's rule, the wavelength corresponding to harmonic degree $n$ is  $\lambda=\frac{2\pi a}{n + \frac{1}{2}}$ {where $a$ is Earth's radius}) the response to surface loads is mostly sensitive to the structure of the lower mantle, where the three variants of VM5a considered here have the same rheological properties. Conversely, for $n=10$ 
{(Figure \ref{fig:earth-loading-1}b)} we see a slightly faster response to the loading for both transient models in the time range between $0.01$ and $1$ kyr. For $n=100$, the transient response of VM5a-BG and VM5a-AD becomes even more enhanced between $0.01$ and $10$ kyr. It is worth to note that, for times less than $\sim 10$ kyr, the two transient versions of VM5a almost yield identical responses, suggesting that an Andrade rheology in the Earth's upper mantle might explain the observed vertical transient deformations in the same way as a Burgers rheology. The differences between the three models are more \marginpar{F\ref{fig:earth-loading-2}} evident in Figure \ref{fig:earth-loading-2}, 
{where we use \alma{} for computing the time derivatives $\dot{h}_n(t)$ (this option was not available in previous versions of the program)}. Compared with the Maxwell model, the transient ones show a significantly larger initial rate of vertical displacement, that differ significantly for Burgers and Andrade. The three rheologies provide 
{comparable responses} only $\sim 0.1$ kyrs after loading. {We shall remark, however, that the incompressiblity approximation employed in \alma{} has a significant impact on the $h_n$ Love Number, as we discussed in Section \ref{sect:benchmark_prem}, so the results shown above must be taken with caution, and a more detailed analysis of the impact of compressibility on the time evolution of LNs would be in order.}

\subsection{Tidal dissipation on the Moon}\label{sec:appl_moon}

The Moon is the extraterrestrial body for which the most detailed information about the internal structure is available. In addition to physical constraints from observations of tidal deformation \citep{williams2014lunar}, seismic experiments deployed during the Apollo missions \citep{nunn2020lunar} provided instrumental recordings of moonquakes which allowed the formulation of a set of progressively refined interior models \citep[see, \textit{e.g.}][]{heffels2021reevaluation}.

\begin{figure}
\begin{center}
\includegraphics[width=10cm]{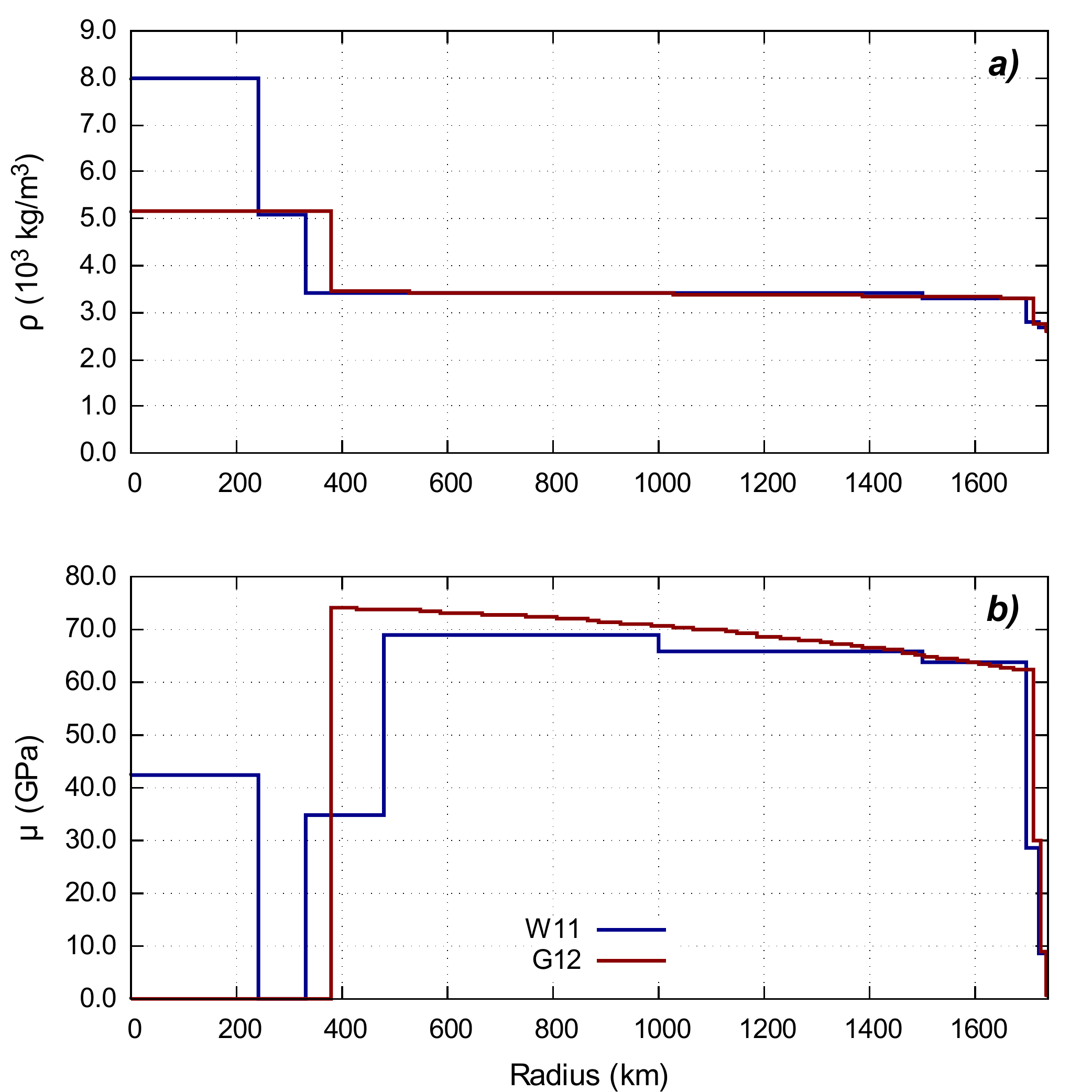}
\end{center}
\caption{Radial profiles of density (a) and rigidity (b) for the Moon models by \citet[]{weber2011moon} (W11, blue) and \citet[]{garcia2011vpremoon,garcia2012erratum} (G12, red). {Models W11 and G12 include $10$ and $71$ homogeneous layers, respectively}.}\label{fig:moon-1}
\end{figure}

In this last numerical experiment, we configured \alma~ to compute tidal LNs for the Moon 
{according to}
the two interior models proposed by \citet[W11 hereafter]{weber2011moon} and \citet[G12 hereafter]{garcia2011vpremoon,garcia2012erratum}. Profiles of density $\rho$ and rigidity $\mu$ for models W11 and G12 are shown \marginpar{F\ref{fig:moon-1}} in Figure~\ref{fig:moon-1}, with the most notable difference being that the former assumes an inner solid core and a fluid outer core, while the latter contains an undifferentiated fluid core. {We emphasize that model G12 includes $70$ rheological layers in the mantle and crust, demonstrating the stability of \alma{} with densely-layered  planetary models}. For both models, we assumed a Maxwell rheology in the crust and the mantle, with a viscosity of $10^{20}$ Pa$\cdot$s. {A more realistic approach has been followed by  \citet{nimmo2012dissipation}, who have modelled the Moon's Love numbers and dissipation adopting an extended Burgers model for the mantle, which also accounts for transient tidal deformations \citep{faul2015transient}. Such rheological 
model is
not incorporated in the current release of \alma{}, but it can be implemented by the user modifying the source code in order to compute the corresponding complex rigidity modulus $\mu(s)$.} 
The fluid core has been modeled as a Newtonian fluid with viscosity $10^4$ Pa$\cdot$s while in the inner core, for model W11, we used a Maxwell rheology with a viscosity of $10^{16}$ Pa$\cdot$s, a value within the estimated ranges for the viscosity of the Earth inner core \citep{buffett1997coreviscosity,dumberry2010inner,koot2011innercore}. Following the lines of \citet{harada2014strong,harada2016deep} and \citet{organowski2020viscoleastic}, we defined a $150$ km thick low-viscosity zone (LVZ) at the base of the mantle and computed the $k_2$ tidal LNs as a function of the LVZ viscosity for a forcing period $T=27.212$ days. 

\begin{figure}
\begin{center}
\includegraphics[width=10cm]{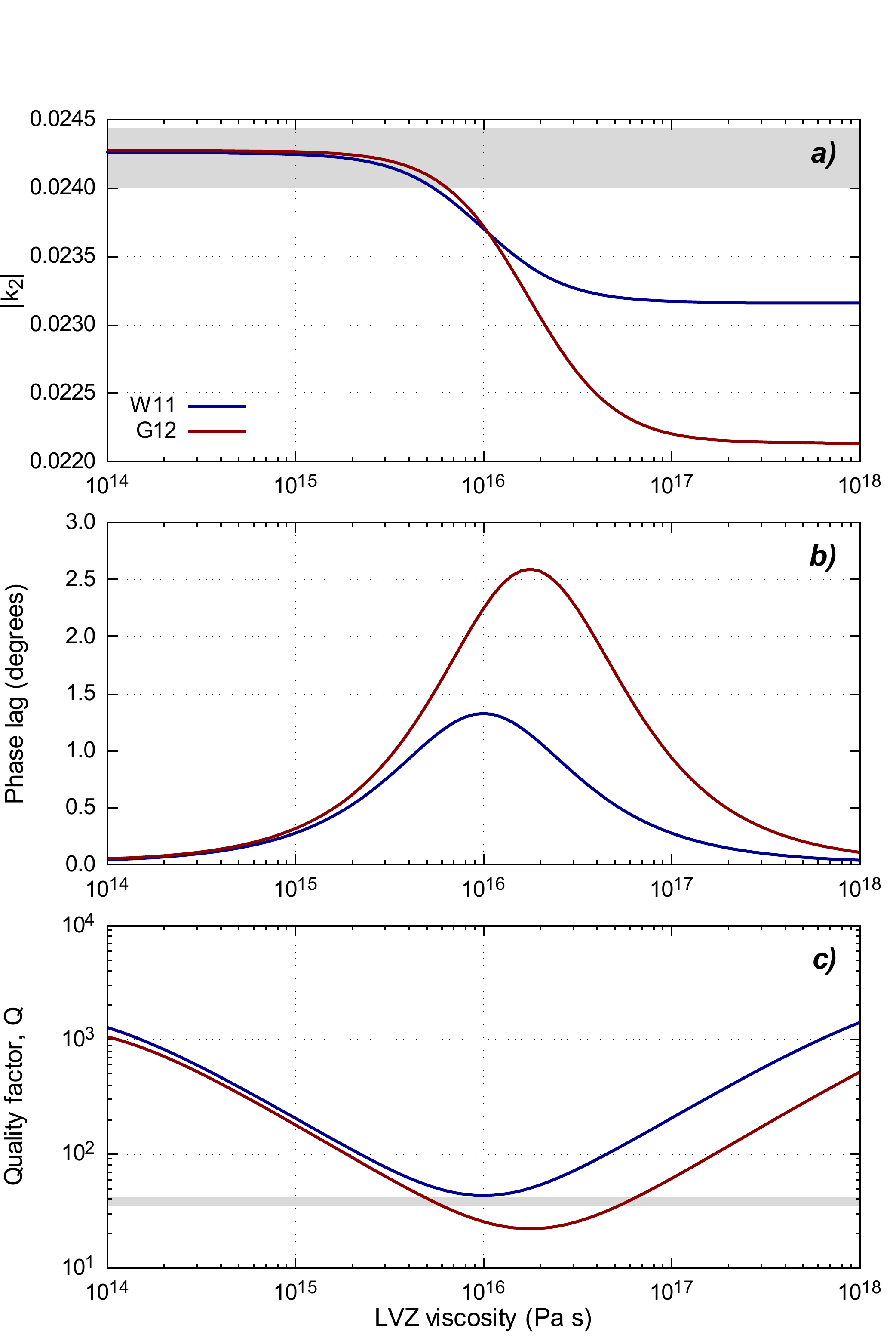}
\end{center}
\caption{Modulus of the tidal Love number $|k_2|$ for the Moon (frame a), phase lag (b) and quality factor (c) as a function of the LVZ viscosity, for a forcing period $T=27.212$ days. Blue and red curves correspond to the Moon models by \citet{weber2011moon} and \citet{garcia2011vpremoon,garcia2012erratum} shown in Figure~\ref{fig:moon-1}. Shaded areas in frames (a) and (c) correspond to the 1-$\sigma$ confidence intervals for measured values of $k_2$ and $Q$ {according to \citet{williams2015moon}.}}\label{fig:moon-2}
\end{figure}

For both W11 and G12 models, Figure~\ref{fig:moon-2} \marginpar{F\ref{fig:moon-2}} shows the dependence on the LVZ viscosity of the $k_2$ tidal LN {(Figure \ref{fig:moon-2}a)}, of its phase lag angle {(\ref{fig:moon-2}b)} and of the quality factor $Q$ {(\ref{fig:moon-2}c)}. With the considered setup, for a LVZ viscosity smaller than $10^{15}$ Pa$\cdot$s the tidal response of the two models is almost coincident, while for higher viscosities model G12 predicts a stronger tidal dissipation. Shaded gray areas in frames 
{(\ref{fig:moon-2}a)} and {(\ref{fig:moon-2}c)} show 1-$\sigma$ confidence intervals for experimental estimates of $k_2$ \citep{williams2014lunar} and $Q$ \citep{williams2015moon}. With both models we obtain values of $k_2$ within the 1-$\sigma$ interval for an LVZ viscosity smaller than about $5\times 10^{15}$ Pa$\cdot$s; interestingly, for that LVZ viscosity the G12 model predicts a quality factor $Q$ within the measured range, while model W11 would require a slightly higher LVZ viscosity ($10^{16}$ Pa$\cdot$s). Of course, a detailed assessment of the ability of the two models to reproduce the observed tidal LNs would be well beyond the scope of this work, and several additional parameters potentially affecting the tidal response (as \textit{e.g.} the LVZ thickness or the core radius) would need to be considered.

\section{Conclusions} \label{sec:conclusions}  

{We have revisited the Post-Widder approach in the context of evaluating viscoelastic Love numbers and their time derivatives for arbitrary planetary models}. Our results are the basis of {a new version of \alma{}}, a user friendly Fortran program that computes the Love numbers of a multi-layered, self-gravitating, spherically symmetric, incompressible planetary model characterized by a linear viscoelastic rheology. \alma{} can be suitably employed to solve a wide range of problems, either involving the surface loading or the tidal response of a rheologically layered planet. By taking advantage of the Post-Widder Laplace inversion method, the evaluation of the time-domain Love numbers is simplified, avoiding some of the limitations of the traditional viscoelastic normal mode approach. Differently from previous implementations \citep{spada2008alma}, \alma{} can evaluate both time-domain and frequency-domain Love numbers, for an extended set of linear viscoelastic 
constitutive equations that also include a transient response, like Burgers or Andrade rheologies. 
Generalized linear rheologies that until now have been utilized in flat geometry like the one characterizing {the extended Burgers model} \citep{ivins2020linear} 
could be possibly implemented as well {modifying the source code, if the corresponding analytical expression of the complex rigidity modulus is available}. 
Furthermore, \alma{} can compute the time-derivatives of the Love numbers, and can deal with step-like and ramp-shaped forcing functions. The resulting Love numbers can be linearly superposed to obtain the planet response to arbitrarily time evolving loads. 
Numerical results from \alma{} have been benchmarked with analytical expressions for a uniform sphere and 
{with a reference set of viscoelastic LNs for an incompressible Earth model \citep{spada2011benchmark}. The well-known limitations of the incompressibility approximation in modeling deformations of large terrestrial bodies have been quantitatively assessed by a comparison between numerical outputs of \alma{} and viscoelastic LNs recently obtained by \citet{Michel-and-Boy-2021} for a realistic, compressible Earth model.
}
The versatility of \alma{} has {then} been demonstrated by a few examples, in which the Love numbers and some associated quantities like the quality factor $Q$, have been evaluated for some multi-layered models of planetary interiors characterized by complex rheological profiles and by densely-layered internal structures.

\section*{Acknowledgments}
{We thank the Associate Editor Gael Choblet
and two anomymous Reviewers for their {very} constructive comments that considerably helped improving {our} manuscript. }
{We have benefited from discussion with all scientists involved in the project ``LDLR -- Lunar tidal Deformation from Earth-based and orbital Laser Ranging'', funded by the French ANR and the German agency DGF.} {We are indebted to Steve Vance, Saikiran Tharimena, Marshall Styczinski and Bruce Bills for encouragement and advice.}
DM is funded by a INGV (Istituto Nazionale di Geofisica e Vulcanologia) 2020-23 ``ricerca libera'' research grant and partly supported by the INGV project Pianeta Dinamico 2021-22 Tema 4 KINDLE (grant no. CUP D53J19000170001), funded by the Italian Ministry of University and Research ``Fondo finalizzato al rilancio degli investimenti delle amministrazioni centrali dello Stato e allo sviluppo del Paese, {Legge} 145/2018''. 
CS is funded by a PhD grant of the French Ministry of Research and Innovation. CS {also} acknowledges the ANR, project number ANR-19-CE31-0026 project LDLR (Lunar tidal Deformation from Earth-based and orbital Laser Ranging). GS is funded by a FFABR (Finanziamento delle 
Attivit\`a Base di Ricerca) grant of MIUR (Ministero dell’Istruzione, dell’Università e della Ricerca) and by a {RFO} research grant of DIFA (Dipartimento di Fisica e Astronomia ``Augusto Righi'') of the Alma Mater Studiorum {Università} di Bologna.

\section*{Data availability.}
Source code of the \alma~ version used to obtain numerical results presented in this work is available as a supplementary material. {The latest version of \alma{} can be downloaded from \texttt{https://github.com/danielemelini/ALMA3}}. The data underlying plots shown in this article are available upon request to the corresponding author. 

\bibliographystyle{gji}
\bibliography{scibib-GIORGIO.bib}


\clearpage
    \begin{table}
        \caption{Complex rigidities $\mu(s)$ for the linear viscoelastic rheologies implemented in \alma. Here, $\mu$ is the elastic rigidity, $\eta$ is the {Newtonian} viscosity, $\mu_2$ and $\eta_2$ are the rigidity and viscosity of the transient element in the bi-viscous Burgers rheology, respectively. In the Andrade rheological law, $\alpha$ is the creep parameter while $\Gamma(x)$  
        {is} the Gamma function.}
    \centering
    \begin{tabular}{c|c}
        \hline 
        Rheological law & Complex rigidity $\mu(s)$\\ 
        \hline \\
        Hooke & $\mu$ \\ \\
        Maxwell  &  $\frac{\displaystyle \mu s}{\displaystyle s + \mu/\eta}$ \\ \\
        Newton  &  $\eta\, s $ \\       \\
        Kelvin  &  $\mu + \eta s$ \\ \\
        Burgers  &  $
        \frac{\displaystyle \mu s  \left( s + \frac{\mu_2}{\eta_2} \right)}{\displaystyle
            s^2 + s \left(\frac{\mu}{\eta} + \frac{\mu+\mu_2}{\eta_2} \right) + \frac{\mu\,\mu_2}{\eta\,\eta_2} }$ \\ \\
        Andrade  &  $\left[ \frac{\displaystyle 1}{\displaystyle \mu} + \frac{\displaystyle 1}{\displaystyle \eta s} + \Gamma(\alpha+1) \frac{\displaystyle 1}{\displaystyle \mu} \left(\frac{\displaystyle \eta s}{\displaystyle \mu}\right)^{-\alpha} \right]^{-1} $ \\ \\
        \hline
    \end{tabular}
    \vspace{8pt}
    \label{tab:rheologies}
\end{table}


\clearpage

\clearpage

\clearpage

\clearpage

\clearpage

\clearpage

\clearpage

\clearpage

\clearpage

\clearpage

\clearpage
\clearpage

\end{document}